\documentclass[aps, prb, floatfix, twocolumn, amsmath, amssymb, superscriptaddress, longbibliography, nobibnotes]{revtex4-2}

\usepackage{graphicx}
\usepackage{dcolumn}
\usepackage{color}
\usepackage{comment}

\usepackage[utf8]{inputenc}
\usepackage[T1]{fontenc}
\usepackage{mathptmx}
\usepackage{etoolbox}

\usepackage{lmodern}
\usepackage[colorlinks,linkcolor=blue]{hyperref}
\usepackage[nameinlink,capitalise]{cleveref}
\usepackage{dsfont}

\graphicspath{{./fig/}}

\newlength{\figwidth}
\makeatletter
\def\@email#1#2{%
 \endgroup
 \patchcmd{\titleblock@produce}
  {\frontmatter@RRAPformat}
  {\frontmatter@RRAPformat{\produce@RRAP{#1\href{mailto:#2}{#2}}}\frontmatter@RRAPformat}
  {}{}
}%
\makeatother

\begin{document}

\title{Three-Dimensional Niobium Coaxial Cavity with $\sim0.1\,$second Lifetime}

\author{Takaaki~Takenaka}
\email{takaaki.takenaka@ntt.com}
\altaffiliation{These authors contributed equally to this work} 
\affiliation{Basic Research Laboratories, NTT, Inc., 3-1 Morinosato-Wakamiya, Atsugi, Kanagawa 243-0198, Japan}

\author{Takayuki~Kubo}
\altaffiliation{These authors contributed equally to this work} 
\affiliation{High Energy Accelerator Research Organization (KEK), Tsukuba, Ibaraki 305-0801, Japan}
\affiliation{The Graduate University for Advanced Studies (Sokendai), Hayama, Kanagawa 240-0193, Japan} 

\author{Imran~Mahboob}
\affiliation{Basic Research Laboratories, NTT, Inc., 3-1 Morinosato-Wakamiya, Atsugi, Kanagawa 243-0198, Japan}

\author{Kosuke~Mizuno}
\altaffiliation[present address: ]{Global Research and Development Center for Business by Quantum-AI Technology (G-QuAT), National Institute of Advanced Industrial Science and Technology (AIST), Tsukuba, Ibaraki 305-8568, Japan}
\affiliation{Basic Research Laboratories, NTT, Inc., 3-1 Morinosato-Wakamiya, Atsugi, Kanagawa 243-0198, Japan}

\author{Hitoshi~Inoue}
\affiliation{High Energy Accelerator Research Organization (KEK), Tsukuba, Ibaraki 305-0801, Japan}

\author{Takayuki~Saeki}
\affiliation{High Energy Accelerator Research Organization (KEK), Tsukuba, Ibaraki 305-0801, Japan}
\affiliation{The Graduate University for Advanced Studies (Sokendai), Hayama, Kanagawa 240-0193, Japan}

\author{Shiro~Saito}
\affiliation{Basic Research Laboratories, NTT, Inc., 3-1 Morinosato-Wakamiya, Atsugi, Kanagawa 243-0198, Japan}


\date{\today}

\begin{abstract}
We report on the internal quality factor of a three-dimensional niobium quarter-wave coaxial cavity, with mid-temperature annealing, exhibiting $Q_{\rm int} \gtrsim 3\times10^9$ at the single-photon level below 20\,mK, which corresponds to an internal photon lifetime of $\tau_{\rm int}\sim90\,\mathrm{ms}$. 
Moreover, $Q_{\rm int}$ of the mid-temperature annealed cavities remains almost unchanged even after several cooldown cycles and air exposure. 
These results suggest that stable low-loss niobium oxides might be formed by mid-temperature annealing on the surface of three-dimensional niobium cavity. 
This surface treatment could be applicable to the fabrication of 2D superconducting circuits and help improve the lifetime of Nb-based superconducting qubits.
\end{abstract}

\maketitle

\section{INTRODUCTION}

Three-dimensional (3D) superconducting cavities with high internal quality factor ($Q_{\rm int}$) are gaining attention for quantum information processing and sensing.
The well-controlled electromagnetic environment of a 3D cavity enables the realization of a 3D circuit quantum electrodynamics (cQED) architecture with desired coupling between the superconducting qubit and the cavity, while maintaining the long lifetime of the cavity~\cite{Paik_PRL2011,Reagor_PRB2016}. 
This architecture is a promising platform for quantum information processing using the large Hilbert space of the resonator~\cite{Ofek_Nature2016,Gao_Nature2019,Sivak_Nature2023GKP,Ni_Nature2023Bin11}.
Recently, high-$Q$ 3D cavities have also been utilized to measure dielectric losses of the substrates for superconducting qubits~\cite{Checchin_PhysRevApplied2022,Read_PhysRevApplied2023}, and in the search for dark photon dark matter~\cite{Romanenko_PRL2023, Agrawal_PRL2024, Nakazono_cQED-Darkmatter2025}.

High-purity aluminum is widely used in 3D cQED architectures due to its ease of machining and commercial availability, with internal quality factors $Q_{\rm int}$ reaching $\sim 1 \times 10^8$~\cite{Reagor_PRB2016,Kudra_APL2020}.
Niobium is extensively employed in superconducting radio-frequency (SRF) cavities for particle accelerators, 
where internal quality factors exceeding $1\times10^{11}$ have been demonstrated~\cite{Kubo:IPAC2014-WEPRI022, Romanenko_PRL2017, Romanenko_PRAppl2020}, which motivates us to apply SRF-based knowledge to cQED devices.
However, SRF cavities operate under extremely high radio-frequency (RF) fields, typically with electric fields $E > \mathcal{O}(10)\,{\rm MV/m}$ at 1.4\,K~\cite{HshGL, HshModerate, HshDirty},
and thus surface treatments have been specifically developed to mitigate issues arising under such intense electromagnetic conditions~\cite{Padamsee},
including electron field emission~\cite{FE, Cenni_FE}, degradation of $Q$ due to hydrogen contamination near the surface~\cite{hydride} (commonly referred to as $Q$ disease~\cite{Qdesease, Qdesease_Saito}), 
RF breakdown of superconductivity~\cite{KyotoCamera, Ge_2011}, and magnetic flux trapping during cooldown~\cite{Romanenko_APL2014, PhysRevAccelBeams.19.082001}.
In contrast, cQED devices operate under millikelvin temperatures and low-photon number conditions ($\bar{n} < 10^{4}$), where losses due to two-level systems (TLSs) become one of the limiting mechanisms for $Q_{\rm int}$. 
Therefore, surface treatments developed for niobium SRF cavities cannot be directly translated to cQED devices, and novel surface treatments are necessary. 
 
Recent studies have reported that $Q_{\rm int}$ at low electric fields can exceed $10^9$ by optimizing the cavity geometry~\cite{Milul_PRXQuantum2023} or by applying water-buffered oxide etching at temperatures below $10\,^{\circ}\mathrm{C}$ to remove residual fluorine~\cite{Oriani_arXiv2024}. 
However, $Q_{\rm int}$ tends to degrade after air exposure or repeated thermal cycling, likely due to the growth of native oxides or NbH$_x$ formation from surface hydrogen.
Vacuum storage~\cite{Romanenko_PRAppl2020, Posen_2020} and rapid cooldown protocols to suppress hydrogen migration~\cite{Qdesease} can mitigate these effects, but such procedures are often incompatible with the operational constraints of quantum devices and their measurement environments.
On the other hand, annealing niobium under vacuum is simple and yields various improvements in SRF cavities;
annealing at temperatures above $700\,^\circ\mathrm{C}$ can effectively remove hydrogen on the surface~\cite{Qdesease_Saito}, 
and additional mid-temperature annealing has achieved remarkably low surface resistance of $0.2\,{\rm n\Omega}$~\cite{Ito_PTEP2021}.
Therefore, vacuum annealing of niobium could ensure a stable and high-quality surface even under low microwave field conditions, which is crucial for realizing the full potential of niobium in quantum applications. 

In this paper, we investigate optimal surface treatment procedures for niobium cavities to achieve high $Q_{\rm int}$ at millikelvin temperatures and low-fields.
Specifically, we explore the appropriate sequence and combination of surface treatment techniques developed in the SRF community, such as heat treatments at various temperatures.
For instance, high-temperature baking at 900\,$^\circ\mathrm{C}$ after buffered chemical polishing (BCP) can improve $Q_{\rm int}$, although this procedure is not sufficient to prevent degradation over multiple cooldown cycles.
In contrast, after subsequent mid-temperature annealing inspired by Ref.~\cite{Ito_PTEP2021}, we find $Q_{\rm int}$ reaching $\sim3.0\times10^9$ at the single-photon level, twice the highest $Q_{\rm int}$ of $\sim1.5\times10^9$ previously reported for coaxial cavities.
This performance remains nearly unchanged even after several cooldown cycles with 10 hours of air exposure.
These results are consistent with a reduced fraction of $\mathrm{Nb_2O_5}$ on the surface, as revealed by X-ray photoelectron spectroscopy (XPS) measurements.
Notabley, this treatment process does not require SRF-specific procedures such as electropolishing or high-pressure rinsing, and is compatible with Nb-based superconducting qubit fabrication processes.

\section{METHODS}

The quarter-wave coaxial stub cavities (Fig.~\ref{fig1}(a) and \ref{fig1}(b)) are machined from a block of high-conductivity Nb with a residual resistivity ratio (RRR) of $\geq 300$ and is procured from Tokyo Denkai Co., Ltd. 
The height of the center stub is $h=12.0\,\mathrm{mm}$, which yields the $\lambda$/4 mode with resonance frequency $f_r\sim5.5\,$GHz. 
The radii of the outer wall and the center post are $r_{\rm out}=5.4\,\mathrm{mm}$ and $r_{\rm in}=2.0\,\mathrm{mm}$, respectively, 
and are designed to minimize the conductor losses of the $\lambda$/4 mode after $100\text{--}150\,\mu\mathrm{m}$ of surface etching.
The center stub was fabricated in two ways: either by machining alone (A-series), or by machining followed by finished with electrical discharge machining (EDM) (B-series).
We fabricated cavities using different machining methods because surface roughness can influence $Q_{\rm int}$.
However, after applying appropriate surface treatments, no substantial difference in $Q_{\rm int}$ was observed between them.

The cavities were first tested in their after-machined condition (no additiional surface treatment) and subsequently subjected to a series of surface treatments at the Cavity Fabrication Facility~\cite{CFF} of the High Energy Accelerator Research Organization. 
Our approach followed the fundamental steps of standard procedures established for superconducting cavities used in particle accelerators, with high-field-specific preparations being deliberately omitted.
The simplified version of the SRF surface treatment protocol used in this study is as follows:
\begin{enumerate}
    \item Removal of approximately $100\,\mu\mathrm{m}$ of material from the inner surface using buffered chemical polishing (BCP) at room temperature ($\gtrsim 20^\circ\mathrm{C}$).
    \item Baking the cavity in a vacuum furnace at $900\,^\circ\mathrm{C}$ for 3 hours (high-T bake).
    \item Light BCP ("BCP flush") to remove a few additional micrometers of material.
    \item Additional treatment for surface optimization
\end{enumerate}
The meaning and purpose of each step is briefly reviewed in Appendix~\ref{appendix_surface}, and the estimation of removal depth by BCP is described in Appendix~\ref{appendix_BCPdepth}. 
In step 4, which is a purpose-specific step, 
various treatment options exist in SRF protocols, ranging from those optimized for high accelerating fields to those aimed at minimizing surface resistance for high-$Q$ performance. 
Reference~\cite{Ito_PTEP2021} suggests that a mid-temperature anneal at $T \simeq 600\,^{\circ}\mathrm{C}$ significantly reduces the residual surface resistance. 
In Ref.~\cite{Kalboussi_PhysRevApplied2025}, consistent results have recently reported in the low-fields where losses arising from TLS are non-negligible, therefore mid-temperature annealing might be effective for quantum applications.
We examine and compare the following three options for this step:  
(4a) no additional treatment,  
(4b) annealing at $460\,^\circ\mathrm{C}$ for 3 hours, and  
(4c) annealing at $600\,^\circ\mathrm{C}$ for 3 hours.

After completing the surface treatment, the cavities were set for a reflection geometry ($S_{11}$) and mounted in a dilution refrigerator at a base temperature below $20\,\mathrm{mK}$.
Cavity performance is evaluated using both frequency-domain and time-domain measurements.
For the frequency-domain measurement, we use a standard vector network analyzer (VNA) with an IF bandwidth of $\,\sim1\,$Hz. 
The time-domain measurement is carried out with Quantum Machines' OPX+ and Octave system.
The experimental cryogenic setup employed in this study is shown in Fig.~\ref{fig1}(d).

\section{$Q_{\rm int}$ at base temperatures}

\begin{figure*}[tb!]
  \centering
  \includegraphics[clip, width=\linewidth]{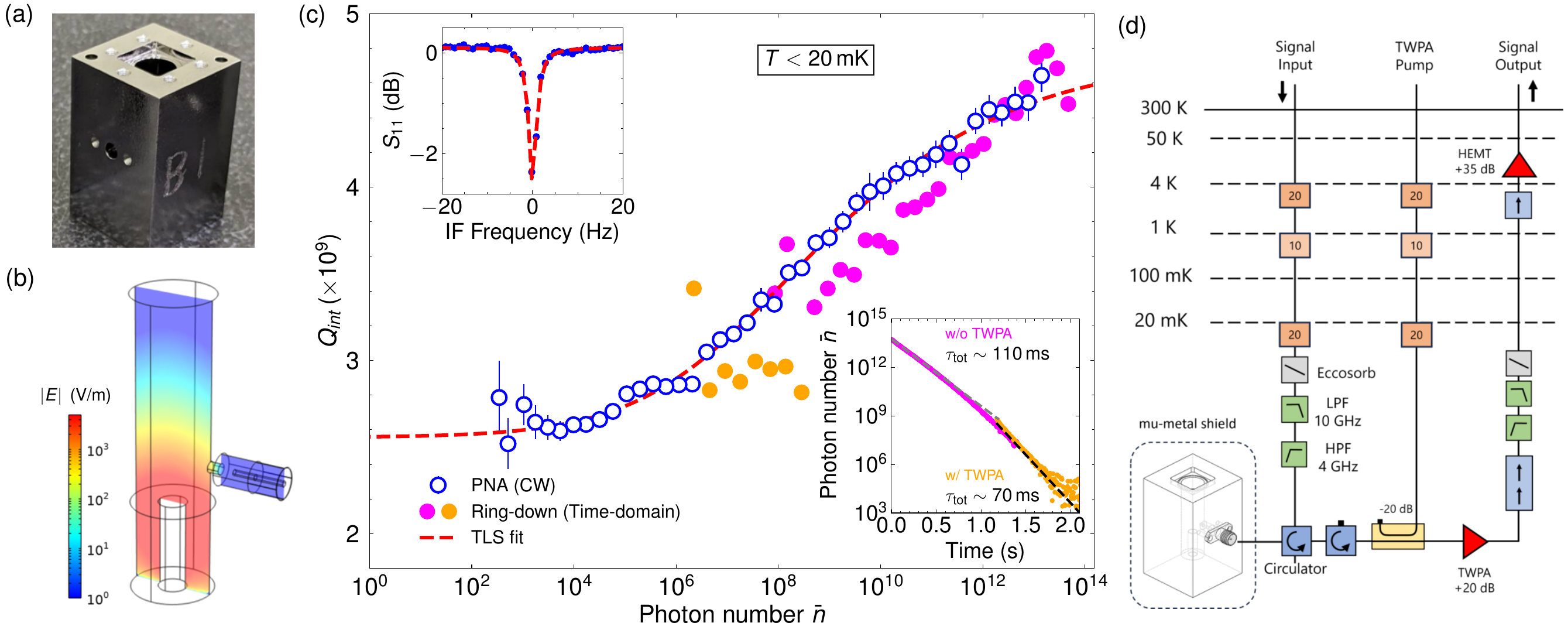}
  \caption{
    (a) A photo of the niobium cavity used in this study. 
    (b) The simulated electric field distribution for the $\lambda$/4 mode of the coaxial cavity with input voltage $V_{\rm in}=0.1\,\mathrm{V}$ from the extetnal coupling port. 
    (c) Characterization of the internal quality factor $Q_{\rm int}$ of cavity B2 after a mid-temperature anneal for 3 hours at 600\,$^\circ\mathrm{C}$.
    The internal quality factor as a function of average photon number $\bar{n}$ in the cavity is fitted using the standard TLS model (Eq.(1)) and is shown in the red dashed line with 
    $F\delta^0_{\rm TLS} = 1.82\times10^{-10}$, $n_c = 2.0\times10^6$, $\beta=0.34$, and $Q_{\rm res}=4.77\times10^{9}$. 
    $Q_{\rm int}$ (points) is extracted from both $S_{11}$ reflection spectra (upper inset) and ring-down measurements (lower inset). 
    Ring-down measurements are performed both with (orange) and without (magenta) a Traveling Wave Parametric Amplifier (TWPA) since the quantum-limited amplifier saturates under a high-power signal. 
    Note that this data was obtained in the second cooldown, after exposing the cavity to air for 10 hours, leading insignificant degradation of $Q_{\rm int}$ compared to the first cooldown.
    (d) A schematics of the low-temperature part of the measurement setup employed for this experiment.
  }
  \label{fig1}
\end{figure*}

Fig.~\ref{fig1}(c) summarizes the dependence of $Q_{\rm int}$ on the average photon number $\bar{n}$ for cavity B2, following 100$\,\mu\mathrm{m}$ BCP, 900\,$^\circ\mathrm{C}$ baking for 3 hours, BCP flush, 600\,$^\circ\mathrm{C}$ annealing for 3 hours, and 10 hours of air exposure.
The upper inset of Fig.~\ref{fig1}(c) displays the magnitude of the reflection coefficient $S_{11}$. 
The experimental data are reproduced by a circle fit (dashed line), from which $Q_{\rm int}\sim 3.0\times10^{9}$ and $Q_{\rm ext}\sim 17\times10^{9}$ are extracted~\cite{Probst_RSI2015}.
The lower inset of Fig.~\ref{fig1}(c) shows the ring-down measurement, yielding the total cavity decay time of approximately 110\,ms in the high-photon regime (magenta) and 70\,ms at low-photon numbers (orange).
The internal decay time can then be extracted, using $\tau_{\rm tot}^{-1} = \tau_{\rm ext}^{-1} + \tau_{\rm int}^{-1}$ and $\tau_{\rm ext} = Q_{\rm ext} / (2 \pi f_r)$, resulting in $\tau_{\rm int, high}\sim141\,$ms and $\tau_{\rm int, low}\sim80\,$ms, respectively.
The two datasets show agreement across a wide range of photon numbers, demonstrating the reliability and reproducibility of both measurement results.

The average photon number dependence of $Q_{\rm int}$ in the low-temperature regime $T<1.2\,\mathrm{K}$ is described by the standard TLS model~\cite{Gao_APL2008, Wang_APL2009, Muller_RevProgPhys2019, Romanenko_PRAppl2020, McRae_RSI2020, Lei_APL2020, Altoe_PRXQuantum2022}
\begin{equation}
  \frac{1}{Q_{\rm int}(\bar{n})} = F \delta^0_{\rm TLS} \frac{\tanh\left( \frac{h f_r}{2k T} \right) }{\sqrt{1 + \left( \frac{\bar{n}}{n_c} \right)^{\beta}}} + \frac{1}{Q_{\rm res}}, \label{TLSmodel-P}
\end{equation}
where $F = t_{ox} \int_S |E^2| dS / \int_V \epsilon_r |E^2| dV$ is the geometric factor of TLSs, $\delta^0_{\rm TLS}$ is the loss tangent of the oxide layer with $T\rightarrow 0$ at the single-photon level (effectively proportional to TLS density), 
$k$ is the Boltzmann constant, $n_c$ is the critical photon number for TLS saturation, $\beta$ is a fitting parameter which reflects the geometry of the cavity,
and $Q_{\rm res}$ is the residual quality factor which arises from magnetic defects or other non-TLS-related losses~\cite{Romanenko_APL2014,PhysRevAccelBeams.19.082001, Gurevich_2017, Kubo_2022}.
We simulate the surface electric participation ratio $S_e = \int_S |E|^2 dS / \int_V |E|^2 dV \sim 650$ by using COMSOL Multiphysics,
and then obtain $F \sim 1.07\times 10^{-7}$ assuming the thickness and relative permittivity of the niobium oxide layer are $t_{ox}=5\,\mathrm{nm}$ and $\epsilon_r = 30$~\cite{Kalboussi_PhysRevApplied2025}, respectively.
The red dashed line in Fig.~\ref{fig1}(c) represents the fit from Eq.~\eqref{TLSmodel-P}, indicating $Q_{\rm int}$ is limited by TLS losses for photon numbers below $n_c$ of $\sim 10^6$, and it approaches the value at single-photon level $\sim 3\times10^9$ (Fig~\ref{fig1}(c)).

\section{Cavity's Temperature dependence}
\begin{figure}[tb!]
  \centering
  \includegraphics[clip, width=0.95\linewidth]{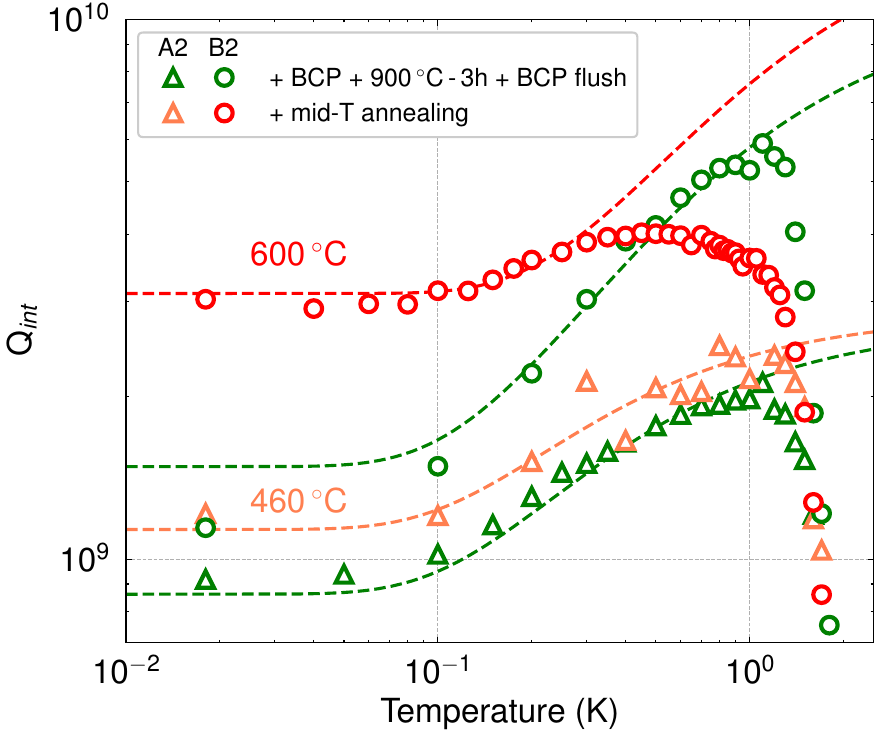}
  \caption{
    Temperature dependence of the internal quality factor $Q_{\rm int}$ after each surface treatment. 
    Dashed lines are fits to the temperature-dependent TLS model [Eq.~\eqref{TLSmodel-T}] using low-power data.
    Results are shown for cavities A2 and B2 after : 100$\,\mu\mathrm{m}$ BCP + 900\,$^\circ\mathrm{C}$ for 3 hours + BCP flush (circles); 
    and subsequent mid-temperature annealing at 460\,$^\circ\mathrm{C}$ or 600\,$^\circ\mathrm{C}$ for 3 hours (triangles). 
    These measurements were performed at an average photon number of $\bar{n}\sim10^4$.
  }
  \label{fig Q-T}
\end{figure}

The evolution of $Q_{\rm int} (T)$ after each treatment process is shown in Figure~\ref{fig Q-T} and Appendix~\ref{appendix_Alltrace}.
Without surface treatment, $Q_{\rm int}$ is less than $1\times10^{8}$ for all cavities. 
The cavity fabricated by machining (cavity A2) exhibits an order of magnitude higher $Q_{\rm int}$ than that fabricated by EDM (cavity B2),
and their temperature dependence differs significantly (blue symbols in Fig.~\ref{fig Q-T, all trace}).
This behavior most likely originates from the formation of a low-$T_c \sim 1\,\mathrm{K}$ material on the cavity surface from EDM.

After high-temperature baking and subsequent BCP flush, all cavities reach $Q_{\rm int}\sim 1 \times10^{9}$ at base temperature ($T<20\,\mathrm{mK}$) and exhibit a peak around $T\sim1.2\,\mathrm{K}$ (green symbols in Fig.~\ref{fig Q-T}).
The temperature dependence of $Q_{\rm int}$ with low input power ($\bar{n}\sim 1\times10^4$) is fitted by the temperature-dependent TLS model~\cite{Romanenko_PRAppl2020,Oriani_arXiv2024,Lei_APL2020} 
\begin{equation}
  \frac{1}{Q_{\rm int}(T)} = F \delta^0_{\rm TLS} \mathrm{tanh}\left(\frac{h f_r}{2k T} \right) + \frac{1}{Q_{\rm int,0}}. \label{TLSmodel-T}
\end{equation}
$Q_{\rm int}(T)$ curves below 1.0\,K are well reproduced using the same $F\delta_{\rm TLS}^0$ obtained from the power-dependent TLS model [Eq.~\eqref{TLSmodel-P}],
and is typically $0.5 \textrm{--} 1.0 \times 10^{-9}$ (see Appendix~\ref{appendix_Alltrace} for details) thus indicating that the same TLSs govern the low temperature behavior.

Subsequent annealing at 460\,$^\circ\mathrm{C}$ for 3 hours improves $Q_{\rm int}$ by a factor of 1.5, whereas 600\,$^\circ\mathrm{C}$ annealing leads to a further improvement by a factor of 3 (red and orange symbols in Fig.~\ref{fig Q-T}). 
$F\delta_{\rm TLS}^0$ is also reduced by the mid-temperature annealing, with values of $\sim4.9\times10^{-10}$ for 460\,$^\circ\mathrm{C}$ annealing and $\sim2.1\times10^{-10}$ for 600\,$^\circ\mathrm{C}$ annealing.
The difference between the 460\,$^\circ\mathrm{C}$ and 600\,$^\circ\mathrm{C}$ annealing is also reflected in the temperature dependence of $Q_{\rm int}$. 
The 460\,$^\circ\mathrm{C}$-annealed cavity exhibits a peak in $Q_{\rm int}$ around 1.2\,K, similar to the cavities without mid-temperature annealing as well as SRF cavities~\cite{Romanenko_PRAppl2020}.
In contrast, the 600\,$^\circ\mathrm{C}$-annealed cavity shows its maximum $Q_{\rm int}$ around 0.3\,K, a behavior previously reported only in the water-buffered chemically etched cavities~\cite{Oriani_arXiv2024}.
The difference will be discussed later.

\begin{figure}[tb!]
  \centering
  \includegraphics[clip, width=0.95\linewidth]{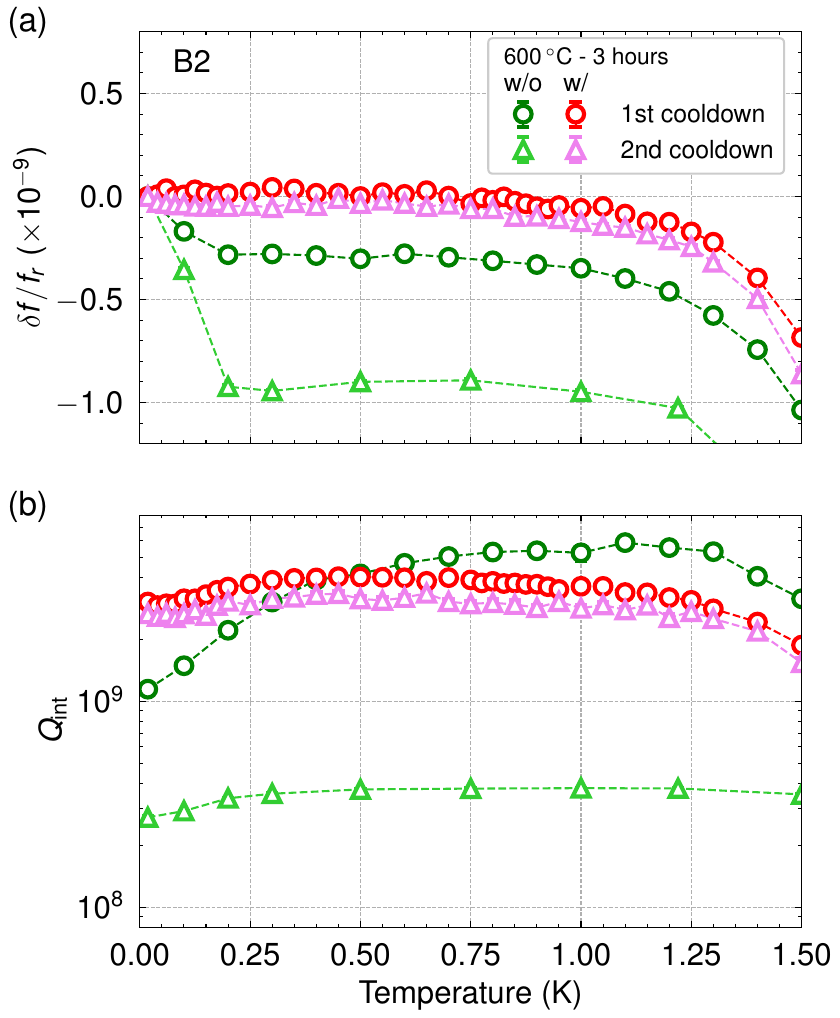}
  \caption{
    Temperature dependence of the normalized frequency shift $\delta f / f_r$ (a) and internal quality factor $Q_{\rm int}$ (b) for cavity B2 measured across multiple cooldown cycles, before and after 600\,$^\circ\mathrm{C}$ mid-temperature annealing.
  }
  \label{fig f-T}
\end{figure}

The low-TLS-loss property of the niobium cavity surface after mid-temperature annealing can also be confirmed through observations of resonance frequency shifts as a function of temperature.
At low temperatures, TLS-induced resonance frequency shifts can be described by the following expression~\cite{McRae_RSI2020, Heidler_PRAppl2021,Altoe_PRXQuantum2022},
\begin{gather}
 \frac{\delta f}{f_r}  = \frac{F\delta^0_{\rm TLS}}{\pi}\left[\mathrm{Re}\Psi\left(\frac{1}{2} + \frac{1}{2\pi i}\frac{hf_r}{kT}\right) - \log \left(\frac{hf_r}{2\pi k T}\right)\right], \label{TLSmodel-f}
\end{gather}
where $\mathrm{Re}\Psi$ denotes the real part of the complex digamma function. 
Eq.~\eqref{TLSmodel-f} gives a frequency minimum around $T \sim h f_r / k$, which corresponds to about 120$\,\mathrm{mK}$ in our $5.5\,$GHz cavities.
Figure~\ref{fig f-T} shows the evolution of normalized frequency shift ${\delta f}/{f_r}$ and $Q_{\rm int}$ for cavity B2 with cooldown cycles and air exposures.
Before mid-temperature annealing, a clear frequency shift from base temperature to $120\text{--}200$\,mK indicates the presence of a non-negligible density of TLSs in the oxide layer on the cavity surface.
In the second cooldown, the frequency shift became more pronounced, and $Q_{\rm int}$ decreases to approximately one-fifth of that in the first cooldown, 
suggesting an increase in $F\delta_{\rm TLS}^0$ and the associated rise in total loss originating from the change in the TLS density.
On the other hand, after 600\,$^\circ\mathrm{C}$ mid-temperature annealing, the frequency shift is dramatically suppressed when $T<1.0\,\mathrm{K}$ in conjunction with an increased $Q_{\rm int}$ thus indicating a reduction in TLS density.

\begin{figure}[tb!]
  \centering
  \includegraphics[clip, width=0.95\linewidth]{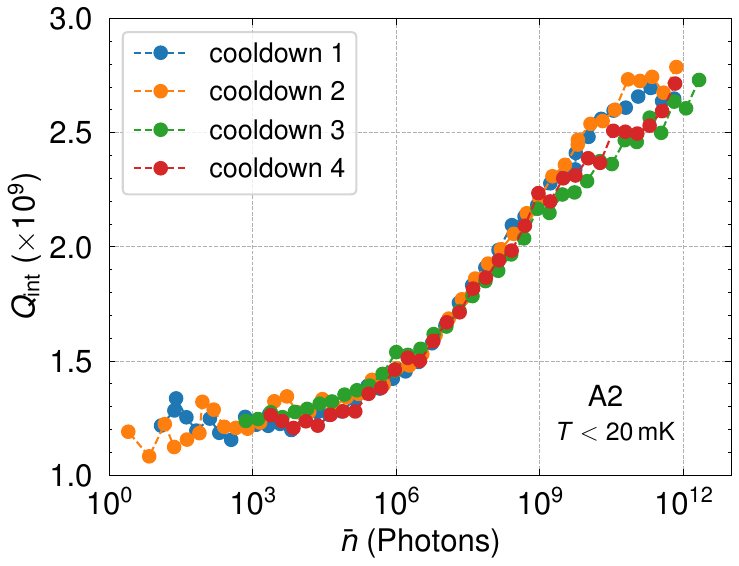}
  \caption{
    Comparison of $Q_{\rm int}$ as a function of average photon number $\bar{n}$ in cavity A2 after 460\,$^\circ\mathrm{C}$ annealing over multiple cooldown cycles.
    The cavity was exposed to air for approximately 2 hours between each cooldown cycles, and it confirms the stability of the cavity performance against air exposure. 
    $Q_{\rm int}$ is extracted from $S_{11}$ spectral measurements with the VNA.
  }
  \label{fig A2-repeat}
\end{figure}

Remarkably, this improved behavior remains nearly unchanged even after 10 hours of air exposure followed by a cooldown, suggesting that the TLS density has not only significantly reduced but it has also been stabilized by the 600\,$^\circ\mathrm{C}$ mid-temperature annealing process.
The 460\,$^\circ\mathrm{C}$ annealed cavity A2 also maintains a high-$Q$ value over multiple cooldown cycles, and its power dependence remains unchanged, as shown in Fig.~\ref{fig A2-repeat}.
In typical niobium cavities for quantum applications that only undergo bulk BCP, $Q_{\rm int}$ drops after several cooldown cycles and the resultant temperature dependence indicates the occurrence of hydrogen $Q$ disease~\cite{Oriani_arXiv2024,Milul_PRXQuantum2023}.
Indeed, our 900\,$^\circ\mathrm{C}$ baked cavities also show similar degradation in $Q_{\rm int}$ within a few cooldown cycles.
This contrasting behavior suggests that mid-temperature annealing plays a significant role in stabilizing the niobium cavities against air exposure and repeated cooldown cycles.

\section{DISCUSSION}

Our cavities, after 900\,$^\circ\mathrm{C}$ baking for 3 hours followed by a BCP flush, exhibited a $2\text{--}3$-fold improvement in $Q_{\rm int}$ compared to previous coaxial cavities treated with conventional BCP alone~\cite{Heidler_PRAppl2021,Oriani_arXiv2024}.
The key difference between our method and previous studies lies in the duration of niobium surface exposure to the BCP solution.
Previously, BCP was typically performed for approximately two hours, and such prolonged exposure to conventional BCP solutions is reported to result in fluoride residue near the surface~\cite{Oriani_arXiv2024} and to promote hydrogen absorption.
In contrast, our method also begins with an initial $100\,\mu\mathrm{m}$ bulk BCP, followed by 900\,$^\circ\mathrm{C}$ baking, which causes contaminants to precipitate onto the surface.
A subsequent short-duration BCP flush then effectively removes these impurities and residues.
Note that fluoride formation should be minimal because the BCP flush lasted only about 30 seconds (see Appendix~\ref{appendix_BCPdepth} for details).
XPS measurements (Appendix~\ref{appendix_XPS}) reveal that a small amount of fluorine remains on the niobium surface after the initial $100\,\mu\mathrm{m}$ bulk BCP, which disappears following high-temperature annealing and BCP flush. 
This result is consistent with the findings of Oriani et al.~\cite{Oriani_arXiv2024}, who demonstrated that H$_2$O-buffered etching provides better performance than conventional ${\rm H_3PO_4}$-buffered etching due to the elimination of fluoride residue on cavity surface.

\begin{figure}[tb!]
  \centering
  \includegraphics[clip, width=0.95\linewidth]{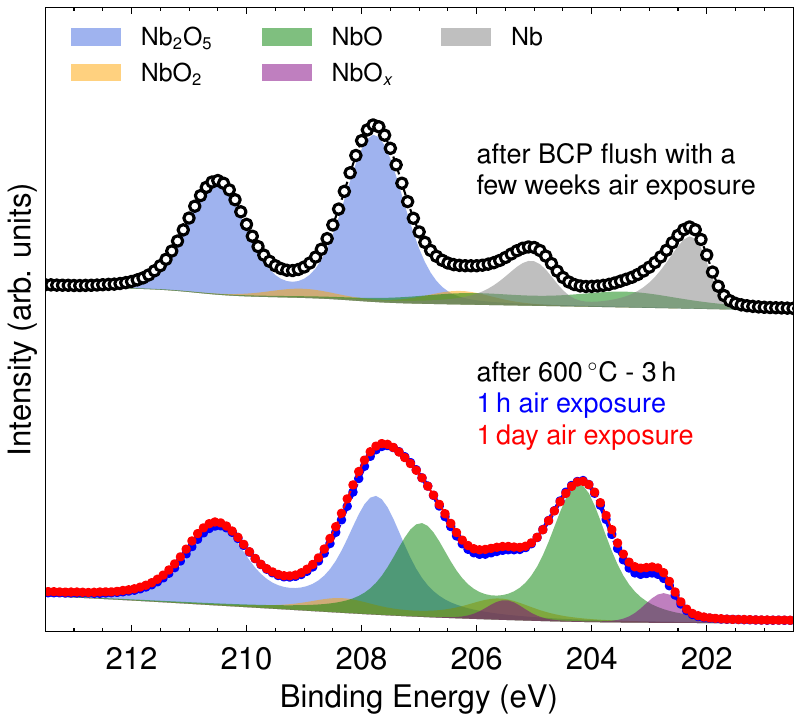}
  \caption{
    XPS spectrum of the Nb $3d$ region after a BCP flush, and subsequent 600\,$^\circ\mathrm{C}$ annealing with 1 hour air and 1 day air exposures. 
    Peaks corresponding to ${\rm Nb_2O_5}$, ${\rm NbO_2}$, NbO, ${\rm NbO_x} (0<x<1)$, and metallic Nb are indicated with filled areas. 
  }
  \label{fig XPS}
\end{figure}

The further improvement in $Q_{\rm int}$ observed after mid-temperature annealing is attributed to modifications to the surface oxides. 
It is well established that vacuum annealing at temperatures above 340\,$^\circ\mathrm{C}$ dissolves ${\rm Nb_2O_5}$, 
which also leads to a reduction in the loss tangent $\delta^0_{\rm TLS}$ in SRF cavities~\cite{Kalboussi_APL2024, Bafia_PhysRevApplied2024,Yu_Vacuum2022}. 
The correlation between the amount of ${\rm Nb_2O_5}$ and the quality factor was also demonstrated in 2D superconducting circuits at low temperatures $\sim10\,\mathrm{mK}$, 
both by controlling the thickness of the ${\rm Nb_2O_5}$ layer~\cite{Verjauw_PhysRevApplied2021} and by selectively removing ${\rm Nb_2O_5}$ from the surface~\cite{Altoe_PRXQuantum2022}. 
Figure~\ref{fig XPS} shows a high-resolution scan of Nb $3d$ region before and after 600\,$^\circ\mathrm{C}$ annealing.
Before applying 600\,$^\circ\mathrm{C}$ annealing, clear peaks corresponding to ${\rm Nb_2O_5}$ are observed around $210\,$eV,
while after annealing, those peaks are reduced and NbO-related peaks around $204\,$eV become more pronounced.
The composition ratio of $\mathrm{Nb_2O_5}$ decreases from $\sim60$\,\% to $\sim40$\,\%, while NbO increases from $\sim10$\,\% to $40$\,\% (see Appendix~\ref{appendix_XPS} for details).
These findings are consistent with the recent study by Kalboussi et al.~\cite{Kalboussi_PhysRevApplied2025}, which demonstrated that 650\,$^\circ\mathrm{C}$ vacuum annealing increased $Q_{\rm int}$ by approximately an order of magnitude at $T\sim1.4\,\mathrm{K}$.

This modification of the surface niobium oxides is consistent with the distinct temperature dependence of $Q_{\rm int}$ observed in the 600\,$^\circ\mathrm{C}$-annealed cavity, as shown in Fig.~\ref{fig Q-T}. 
At low temperatures, the behavior of $Q_{\rm int}(T)$ can be attributed to the interplay between TLS losses and quasiparticle excitations, with the latter typically becoming significant when $T/T_c \gtrsim 0.3$, which corresponds to $T \simeq 0.4\,\mathrm{K}$ for NbO with $T_c \sim 1.4\,\mathrm{K}$. 
The experimental observation that $Q_{\rm int}$ reaches a maximum at $T \sim 0.3\,\mathrm{K}$ after 600\,$^\circ\mathrm{C}$ annealing, 
together with the emergence of a frequency shift at $T \gtrsim 0.5\,\mathrm{K}$, is therefore consistent with the temperature scale at which quasiparticle excitations in NbO become non-negligible.

The substantial presence of NbO near the surface could account for the stability of $Q_{\rm int}$ against air exposure.
It has been suggested that on an unannealed niobium surface, the ${\rm Nb_2O_5}$ layer grows over time, whereas ${\rm NbO_2}$ and ${\rm NbO}$ remain at monolayer thickness~\cite{Verjauw_PhysRevApplied2021}.
In contrast, it has been reported that 300\text{--}600$^\circ\mathrm{C}$ annealing increases the thickness of the NbO layer between ${\rm Nb_2O_5}$ and metallic Nb~\cite{Yu_Vacuum2022,Kalboussi_PhysRevApplied2025},
which can slow oxide growth by increasing the potential barrier for oxygen incorporation.
Our observation --- $Q_{\rm int}$ remains stable under air exposure or thermal cycles after annealing at 460\,$^\circ\mathrm{C}$ or 600\,$^\circ\mathrm{C}$ --- is consistent with these studies.
The presence of a thick NbO layer on the Nb surface might be harmful to SRF applications in which cavities operate at $T\sim1.4\,\mathrm{K}$.
For quantum applications, however, such a layer is unlikely to degrade performance because the devices operate well below the $T_c$ of NbO; moreover, it helps stabilize performance under repeated operations. 
Therefore, although mid-temperature annealing in vacuum is much simpler than the techniques used for SRF cavities, it has the potential to be a more practical approach to achieving improved performance in Nb-based quantum devices.

\section{CONCLUSIONS}

In summary, we have demonstrated that a SRF-inspired surface treatment recipe significantly improves the internal quality factor $Q_{\rm int}$ of niobium coaxial cavities at low temperatures $\sim 20\,\mathrm{mK}$ and at the single-photon level.
All cavities after applying 100$\,\mu\mathrm{m}$ BCP, 900\,$^\circ\mathrm{C}$ baking for 3 hours, and subsequent BCP flush for outermost surface removal exhibit $Q_{\rm int}\sim1\times10^9$.
$Q_{\rm int}$ is then further improved by an additional 600\,$^\circ\mathrm{C}$ annealing for 3 hours, achieving a record-breaking value of $Q_{\rm int}\sim3\times10^9$ in $\lambda/4$ coaxial cavities.
Moreover, no significant degradation is observed in $Q_{\rm int}$ with repeated cooldown cycles or air exposure, which has been a major issue in niobium cavities for quantum applications.
Experimental observations indicates that these improvement are brought by reduced fraction of ${\rm Nb_2O_5}$ while ${\rm NbO}$ increases by mid-temperature annealing.
Our findings suggest that the long-standing knowledge from SRF cavity research is also beneficial for quantum applications, and it could potentially enable the realization of long-lived Nb-based superconducting quantum platforms.

\section{ACKNOWLEDGMENTS}
The authors thank T. Nitta for fruitful discussions.
We also acknowledge T. Matsushima for his work on XPS measurements, and K. Kakuyanagi and H. Toida for technical assistance.
The cavities were manufactured by Ono-Denki Co. and  Top Seiko Co., Ltd.
This work was supported by JST Moonshot R\&D, Grant Number JPMJMS2067. 
The work of T. K. was supported by JSPS KAKENHI Grant number JP25K23386.

\appendix

\section{Surface treatment}\label{appendix_surface}

Step 1 is the bulk BCP, which plays a critical role in removing the damaged surface layer introduced during fabrication. Decades-old experiments~\cite{1993_Kneisel} have shown that cavities with minimal surface removal exhibit surface resistance nearly an order of magnitude higher than those with approximately $100\,\mu\mathrm{m}$ of material removed.
The chemical solution used for surface removal consists of hydrofluoric acid (HF, 46\%), nitric acid (HNO$_3$, 60\%), and phosphoric acid (H$_3$PO$_4$, 85\%) mixed in a 1:1:1 volume ratio. The surface removal process is governed by the following chemical reactions: 
$6\,\mathrm{Nb} + 10\,\mathrm{HNO_3} \rightarrow 3\,\mathrm{Nb_2O_5} + 10\,\mathrm{NO} + 5\,\mathrm{H_2O}$ and 
$\mathrm{Nb_2O_5} + 10\,\mathrm{HF} \rightarrow 2\,\mathrm{NbF_5} + 5\,\mathrm{H_2O}$.
In this process, nitric acid first oxidizes the niobium surface to form a niobium pentoxide layer, which is subsequently dissolved by hydrofluoric acid. Although phosphoric acid does not directly participate in the chemical reactions, it serves to increase the viscosity of the solution and to moderate the overall reaction rate.
Under the standard 1:1:1 mixture at room temperature (approximately $25^\circ\mathrm{C}$), the etching rate is typically around $10\,\mu\mathrm{m}$/min. 
Increasing the proportion of phosphoric acid is known to slow the etching rate and to produce a smoother surface; a 1:1:2 volume ratio is sometimes employed in SRF cavity processing for this reason.
However, the impact of surface roughness on the $Q$ factor in quantum regime remains unclear. Therefore, in this study, we adopt the standard 1:1:1 volume ratio BCP process at room temperature. 
In this study, we achieved an internal quality factor of $Q_{\rm int} \geq 1.0\times10^{9}$ through a combination of the following procedures.
These results suggest that achieving high $Q_{\rm int}$ in the low-power regime does not depend critically on the etching rate, and that room-temperature BCP is sufficient.
After BCP, the cavities were rinsed and ultrasonically cleaned in ultra-pure water for 15 minutes.

Step 2 is also a well-established process following BCP or electropolishing~\cite{1989_Saito}. 
Annealing at temperatures above approximately $700\,^\circ\mathrm{C}$ is essential to degas hydrogen and to prevent the onset of so-called ``hydrogen $Q$ disease,'' a form of $Q$ degradation first identified in the late 1980s~\cite{Qdesease, Qdesease_Saito}.
Hydrogen $Q$ disease occurs when hydrogen introduced during BCP or electropolishing remains in the niobium lattice as a solid solution at room temperature. Upon cooling to around 100\,K, this hydrogen precipitates as niobium hydride.
Niobium hydride, which has a superconducting transition temperature of approximately $T_c \sim 1\,\mathrm{K}$, exhibits significantly higher surface resistance than pure niobium, thereby degrading the $Q$-value. This degradation can be avoided by rapidly cooling the cavity through the temperature region near 100\,K~\cite{Qdesease, Qdesease_Saito}. 
Alternatively, hydrogen can be effectively removed by vacuum annealing at temperatures above $700\,^\circ\mathrm{C}$, which significantly reduces the risk of hydride formation~\cite{Qdesease_Saito}.
This issue becomes more pronounced in higher-purity niobium. Modern commercially available niobium often exhibits residual resistivity ratio (RRR) values exceeding 300, making proper hydrogen degassing treatment even more critical for avoiding hydrogen $Q$ disease.

Another important aspect of annealing is its role in mitigating flux pinning. It is well established that environmental magnetic flux, such as geomagnetic fields or the magnetization of metal components, can become trapped during the cooldown of a cavity, increasing surface resistance and limiting the $Q$ factor,  recognized over half a century ago (e.g., Ref.~\cite{Kneisel_flux}). 
Empirically, annealing at temperatures $\gtrsim 900\,^\circ {\rm C}$ has been shown to significantly reduce the amount of trapped flux, leading to improved $Q$ factors~\cite{Posen_flux}. Furthermore, dynamic studies of vortex behavior in cavity-grade niobium have demonstrated substantially different behavior in annealed niobium~\cite{2021_Ooi, 2025_Ooi}.

Step 3 is the BCP flush, which is intended to remove surface contamination introduced during the annealing process. 
Extensive experience across the SRF community indicates that cavity performance is sensitive to the cleanliness of the vacuum furnace, particularly when a light BCP or electropolishing step after annealing is omitted~\cite{Konomi, 2020_Marc}.
In many SRF processing protocols, light electropolishing of a few micrometers is commonly applied after annealing. 
Electropolishing~\cite{1989_Saito} also provides a smoother surface than BCP, therefore it is the standard process in modern SRF cavity fabrication.
Since the impact of surface roughness on the quality factor $Q$ in the quantum regime remains unclear, therefore we instead employ a BCP flush, which is also used in modern large-scale accelerator projects~\cite{EXFEL}.
After the BCP flush, the cavities were rinsed and ultrasonically cleaned in ultra-pure water for 15 minutes.
In cavity A1 and B1, the BCP flush was repeated twice and the $Q_{\rm int}$ improved after second BCP flush (see Appendix~\ref{appendix_Alltrace}).

Step 4 is a purpose-specific step.
For high-field SRF cavities, a low-temperature vacuum anneal at approximately $100\,^\circ\mathrm{C}$ for 48 hours~\cite{bake} or its modified version~\cite{Bafia_SRF2019} is typically employed at this stage. The low-temperature vacuum bake is known to diffuse oxygen localized near the surface into the bulk~\cite{Ciovati_bake, Lechner_2024}, thereby modifying the mean free path distribution within the London penetration depth. This redistribution is theoretically expected to suppress the surface screening current and enhance the quench field~\cite{Kubo_2017, Wave_2019, Kubo_2021}.
When moderately high fields are required in combination with high $Q$ values, as in certain accelerator applications, a mid-temperature annealing~\cite{Ito_PTEP2021, Posen_2020, Sha_2022} is often employed instead. For SRF cavities operating at $2\,\mathrm{K}$, where the thermally excited quasiparticle contribution is the dominant factor limiting the $Q$ value~\cite{Gurevich_2017, Kubo_Gurevich_2019, Kubo_2022}, the optimum mid-T bake temperature is considered to lie in the range of $300$-$400\,^\circ\mathrm{C}$.
However, in our quantum application, the optimum mid-T annealing temperature does not necessarily coincide with this range. 
Moreover, Reference~\cite{Ito_PTEP2021} suggests that a mid-temperature annealing at approximately $T \gtrsim 600\,^{\circ}\mathrm{C}$ significantly reduces the residual surface resistance and may be effective for quantum applications. 
A more recent study~\cite{Kalboussi_PhysRevApplied2025}, published during the course of our project, also reports the effectiveness of annealing at a similar temperature.

Note that in SRF applications, high-pressure water rinsing~\cite{HPR, HPR_Saito} and subsequent clean-room assembly, both of which are essential to suppress field emission under high-field operation, were also omitted in this study.

\section{Estimation of Material Removal Depth}\label{appendix_BCPdepth}

Here, we describe the procedure used to estimate the amount of material removed during chemical polishing, $\Delta t$. At KEK, two independent methods are employed in parallel, and their results are compared to ensure consistency.
\begin{enumerate}
    \item \textbf{Direct measurement using a thickness gauge.}  
    The cavity wall thickness is measured before and after BCP using a mechanical gauge at predefined reference points.
    \item \textbf{Indirect measurement based on weight loss.}  
    The cavity is weighed before and after BCP using a precision balance, and the removed thickness is estimated from, assuming uniform etching, $\Delta t= \Delta w/(\rho S)$, where $\Delta w$ is the mass loss, $\rho=8.57\,{\rm g/cm^3}$ is the material density of niobium, and $S$ is the surface area ($110.591\,{\rm cm^2}$ for A1 and A2 cavities and $107.98\,{\rm cm^2}$ for B1 and B2 cavities).
\end{enumerate}

The bulk BCP (Step 1) was conducted at $20\,^\circ\mathrm{C}$ for 12 minutes.  
The amount of material removed was estimated as follows.
Using a thickness gauge, the measured removal depths were $112\,\mu\mathrm{m}$, $100\,\mu\mathrm{m}$, $104\,\mu\mathrm{m}$, and $94\,\mu\mathrm{m}$ for cavities A1, A2, B1, and B2, respectively.
In contrast, the corresponding weight losses were $\Delta w = 9.54\,\mathrm{g}$, $9.46\,\mathrm{g}$, $8.90\,\mathrm{g}$, and $8.98\,\mathrm{g}$, which translate to estimated removal depths of $\Delta t = 101\,\mu\mathrm{m}$, $100\,\mu\mathrm{m}$, $96\,\mu\mathrm{m}$, and $97\,\mu\mathrm{m}$ for A1, A2, B1, and B2, respectively, assuming uniform etching and known material density.
These results confirm that approximately $100\,\mu\mathrm{m}$ of material was successfully removed from each cavity.

The BCP flush (Step 3) was conducted at $20\,^\circ\mathrm{C}$ for 30 seconds.  
The amount of material removed was estimated solely by weight loss, since the removal depth was too small to be measured directly using a thickness gauge.
The measured weight losses were $\Delta w = 0.41\,\mathrm{g}$ and $0.39\,\mathrm{g}$, which correspond to estimated removal depths of $\Delta t = 4.3\,\mu\mathrm{m}$ and $4.2\,\mu\mathrm{m}$ for A2 and B2, respectively.

\section{${\rm H_2 O}$ BCP flush}\label{appendix_H2OBCP}
\begin{figure}[tb!]
  \centering
  \includegraphics[clip, width=0.95\linewidth]{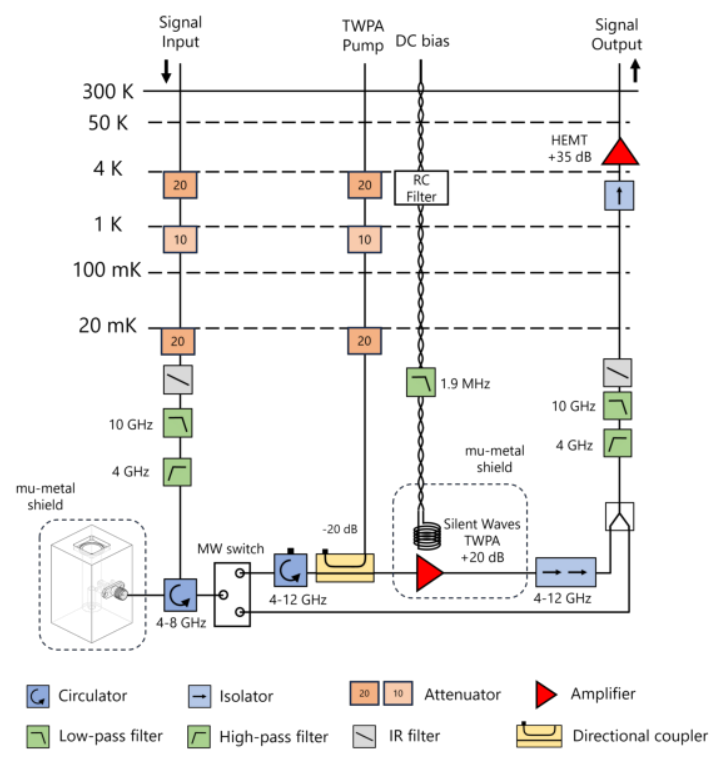}
  \caption{
    Cryogenic measurement setup employed in the experiment. 
    The reflected signal from the sample is routed by a cryogenic RF switch. 
    Under a low-power measurement, the signal is amplified with a TWPA from Silent Waves (The Carthago) by 20\,dB.
    The signal is further amplified by a HEMT amplifier from LNF by 35\,dB.
  }
  \label{fig cryo wiring}
\end{figure}

\begin{figure*}[t!]
  \centering
  \includegraphics[clip, width=\linewidth]{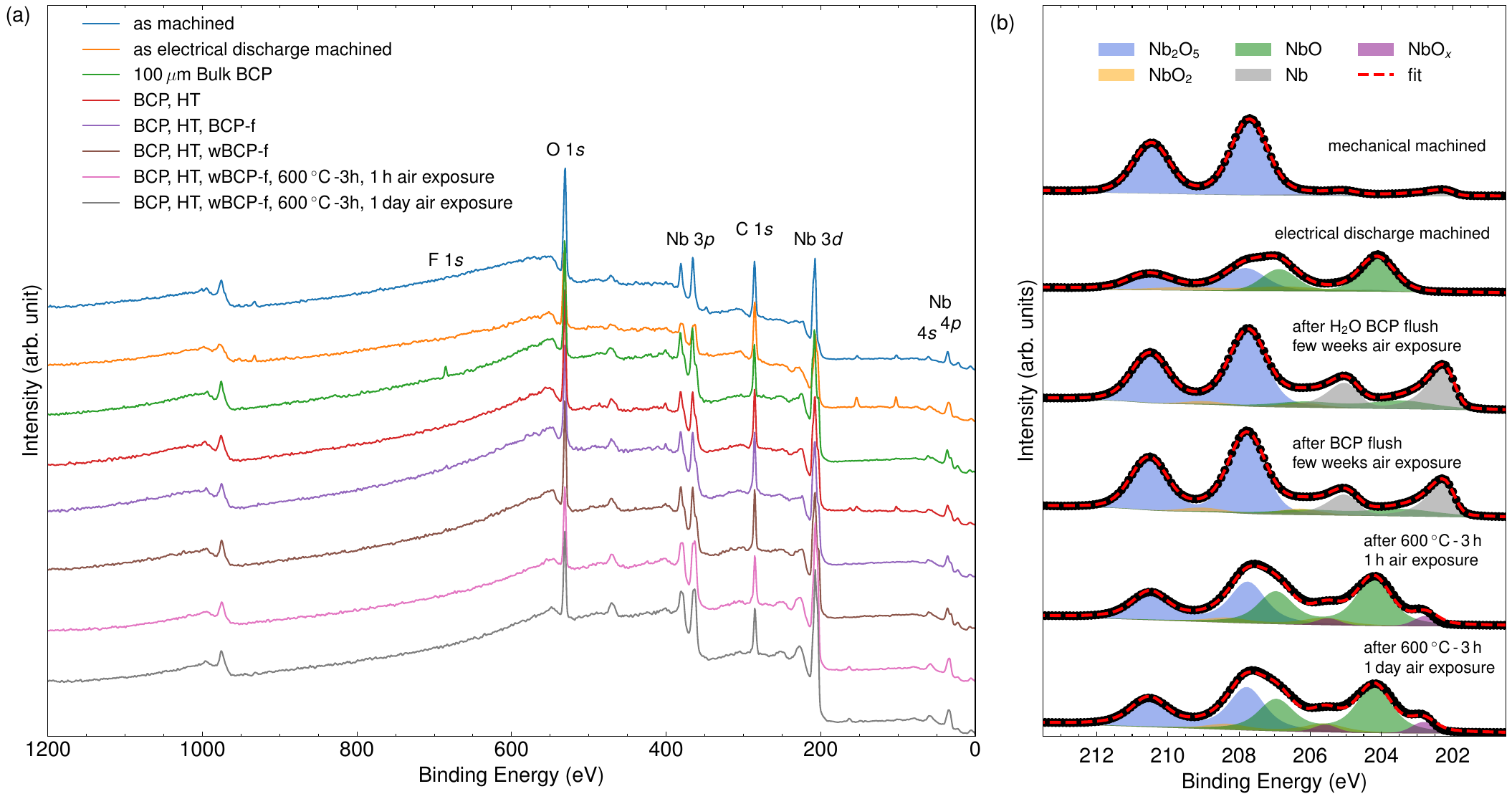}
  \caption{
    (a) Evolution of XPS spectrum before and after various surface treatments. 
    (b) XPS spectrum of Nb $3d$ region for as machined sample and after each procedure (bold black line). 
    Peaks are fitted by considering ${\rm Nb_2O_5}$, ${\rm NbO_2}$, NbO, ${\rm NbO_x} (0<x<1)$, and metallic Nb and they are indicated with filled areas of various colors; 
    the sum of all fitted peaks is then shown as the dashed red line. 
  }
  \label{fig XPS, appendix}
\end{figure*}

We tested a modified BCP solution based on the formulation adopted in Ref.~\cite{Oriani_arXiv2024}, in which water is used as the buffering component instead of ${\rm H_3PO_4}$ in the standard BCP mixture. We refer to this solution as wBCP in the following.

In the first trial, a 1:1:1 volume ratio of HF, HNO$_3$, and ultra-pure water was used at room temperature. Before mixing, the room temperature, which also determines the initial temperature of each chemical, was approximately $16\,^\circ\mathrm{C}$. Upon mixing HNO$_3$ and water, however, the solution temperature rapidly increased to $36\,^\circ\mathrm{C}$. After subsequently adding room-temperature HF and waiting for several minutes, the wBCP solution stabilized at approximately $23\,^\circ\mathrm{C}$.
A wBCP flush was then applied to the A1 and B1 cavities for 25 seconds, followed by rinsing and ultrasonic cleaning in ultra-pure water for 15 minutes.
The removal depths, estimated from weight loss measurements, were $\Delta t = 3.90\,\mu\mathrm{m}$ and $4.01\,\mu\mathrm{m}$ for A1 and B1, respectively.
The measured internal quality factors at $T < 20\,\mathrm{mK}$ were relatively low: $Q_{\rm int} = 4.55 \times 10^8$ for A1 and $6.63 \times 10^8$ for B1 (Table \ref{appendix: tab-value}).

In the second trial, a wBCP solution with a 1:1:2 volume ratio was prepared.  
This time, the container holding the solution was immersed in ice water, and the solution temperature was maintained at $5\,^\circ\mathrm{C}$.  
Taking into account the exponentially slower reaction rate at this low temperature, a wBCP flush was applied to the A1 cavity for 15 minutes, followed by rinsing and ultrasonic cleaning in ultra-pure water for 15 minutes.
The removal depth, estimated from the weight loss measurement, was $\Delta t = 15\,\mu\mathrm{m}$.  
The measured internal quality factor at $T < 20\,\mathrm{mK}$ was $Q_{\rm int} = 1.15 \times 10^9$, which was better than that obtained in the first wBCP trial but was only comparable to the values achieved with the standard BCP flush using ${\rm H_3PO_4}$ as the buffering component.  
For example, the B1 and B2 cavities treated with the standard BCP flush at room temperature exhibited internal quality factors of $Q_{\rm int} = 8.71 \times 10^8$ and $1.14 \times 10^9$, respectively, measured at base temperature $T<20\,\mathrm{mK}$ (Table \ref{appendix: tab-value}).
We did not observe any clear indication that low-temperature wBCP provides superior performance compared to the room-temperature standard BCP process.

\section{Sample preparation and Measurement setup}
\begin{figure*}[tb!]
  \centering
  \includegraphics[clip, width=0.95\linewidth]{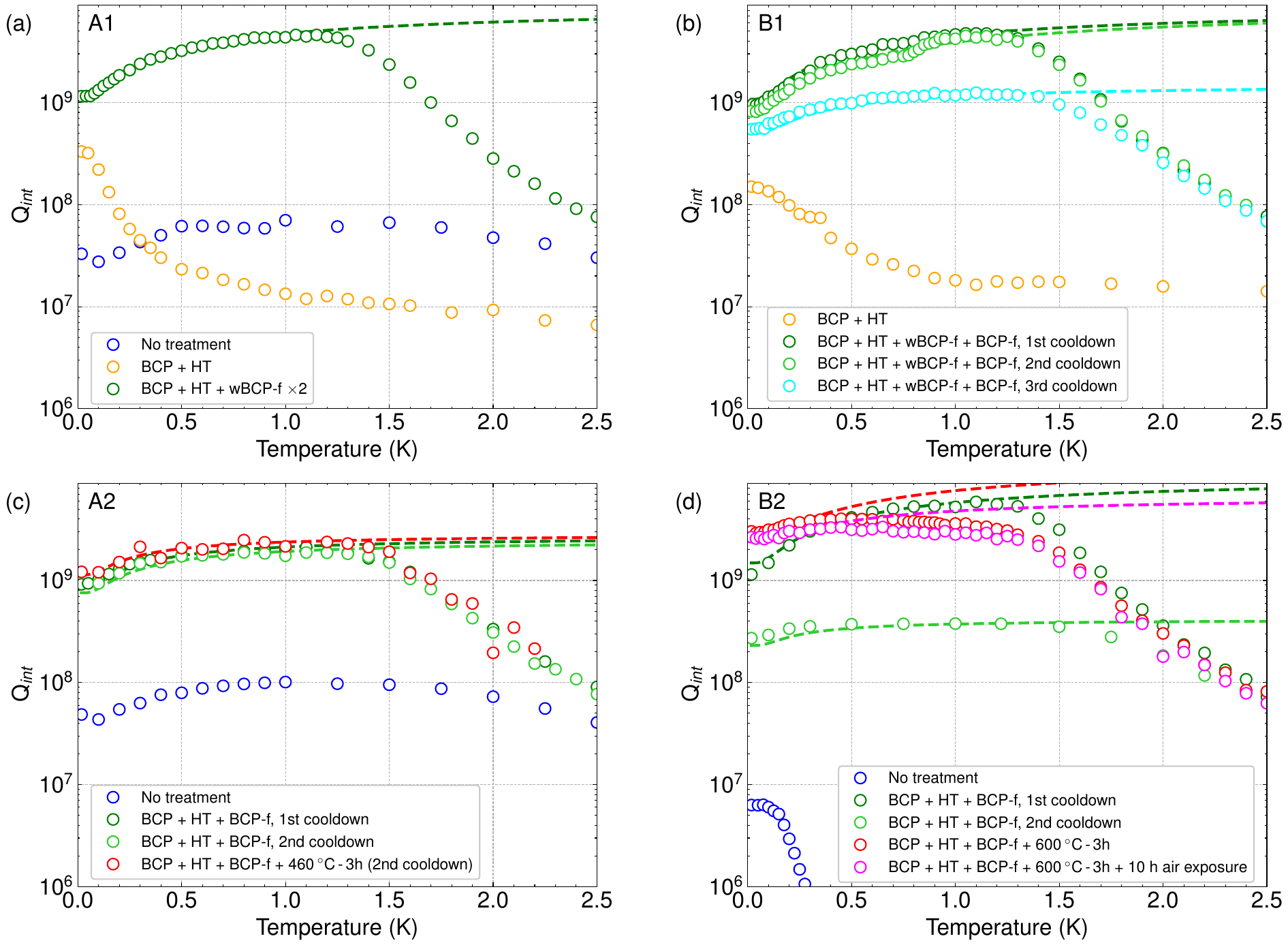}
  \caption{
    Temperature dependence of the $Q_{\rm int}$ for all measured cavities (a) A1, (b) B1, (c) A2, (d) B2.
    (BCP): 100$\,\mu\mathrm{m}$ BCP, (HT): 900\,$^\circ\mathrm{C}$\,-\,3\,h, (BCP-f): BCP flush with standard BCP mixture, (wBCP-f): water-buffered BCP flush.
    Dashed lines are fit for $Q_{\rm int} (T)$ with temperature-dependent TLS model (Eq.~\ref{TLSmodel-T}). 
    These measurements were performed at an average photon number of $\bar{n}\sim10^4$.
  }
  \label{fig Q-T, all trace}
\end{figure*}

After surface treatments, the cavities were stored in air for several days to stabilize the growth of native niobium oxides.
Only the B2 cavity, after 600\,$^\circ\mathrm{C}$ mid-temperature annealing, was sealed in vacuum immediately after removal from the vacuum furnace and opened 30 minutes before installation into the experimental setup.
All cavities underwent multiple cooldown cycles; after warming back to room temperature, they were exposed to air for several hours during the process of exchanging other samples mounted on the dilution refrigerator.

The external quality factor $Q_{\rm ext}$ of the cavity is controlled by adjusting the insertion depth of the SMA connector's center pin, which couples to the cavity for reflection measurements.
$Q_{\rm ext}$ is set to a slightly overcoupled condition ($Q_{\rm ext} \leq Q_{\rm int}$), 
which enables measurements in the quasiparticle (QP) dominant regime $T\geq1.5\,\mathrm{K}$ with high signal-to-noise ratio, where $Q_{\rm int}$ is expected to decrease.
In contrast, the 600\,$^\circ\mathrm{C}$-annealed B2 cavity is measured under both undercoupled ($Q_{\rm ext} > Q_{\rm int}$) and slightly overcoupled conditions.
This is because, under overcoupled conditions, the total cavity decay time $\tau_{\rm tot}$ is dominated by the external decay time $\tau_{\rm ext}$, making it difficult to observe the intrinsic long-lived nature of the high-Q cavity in the time domain.

The top cap of a high-Q coaxial cavity is typically sealed with indium.
In this work, however, measurements were performed either without a cap or with a cap secured only by screws.
This is because the electromagnetic fields of the $\lambda/4$ mode decay to insignificance at the cap position, as the distance from the tip of the center stub is approximately $25\,\mathrm{mm}$. 
Under these measurement configurations, no significant difference in $Q_{\rm int}$ was observed with or without the cap. 
Nonetheless, indium sealing may still provide further improvement in $Q_{\rm int}$.


The detailed cryogenic setup is shown in Fig.~\ref{fig cryo wiring}.
An RF switch (Radiall R577433002) is installed after the circulator, enabling measurement of the reflection coefficient $S_{11}$ of the cavity under both high-power and low-power signals.
For low-power signal measurements, the signal is amplified with a TWPA (Silent Waves, The Carthago). 
The TWPA requires a DC magnetic flux for optimal performance, and the DC current for the local magnet (mounted inside the TWPA package) is supplied by an ADCMT 6166 high-precision DC voltage-current source.
An Agilent E8267C is used to generate the RF pump signal for the TWPA.
Frequency-domain measurements were performed by an Agilent N5232A PNA-L network analyzer.

\section{XPS measurements}\label{appendix_XPS}

To identify the niobium oxides and other residues present on the cavity surface after each treatment step, we executed XPS measurement and analysis for reference samples.
The XPS measurements were carried out using an XPS5700 (PHI) at $45^{\circ}$ with a monochromatic ${\rm Al-K\alpha}$ source ($1486.6\,{\rm eV}$).
The spectra were fitted using LG4X-V2~\cite{LG4X-V2}.
For peak fitting, an asymmetric line shape was employed for metallic niobium, while a symmetric Gaussian-Lorentzian line shape was used for niobium oxides.
An area constraint of $3:2$ was applied to the spin-orbit doublets of Nb\,$3d_{3/2}$ and Nb\,$3d_{5/2}$ levels, and the binding energy separation between them was fixed at 2.75\,eV.
It is assumed that $\mathrm{Nb_2O_5}$ forms on the outermost surface of all samples, and therefore its binding energy was used for peak alignment~\cite{Verjauw_PhysRevApplied2021,Altoe_PRXQuantum2022, Yu_Vacuum2022}.

Figure~\ref{fig XPS, appendix}(a) shows the survey spectrum of the reference niobium sample after each treatment procedure.
The surface after the initial $\sim100\,\mu\mathrm{m}$ bulk BCP shows a clear signature of F $1s$ around $685\,$eV, implying the presence of fluorine bound to niobium~\cite{Oriani_arXiv2024, Prudnikava_SST2024}. 
Figure~\ref{fig XPS, appendix}(b) displays the corresponding XPS spectra of the Nb $3d$ region.
The spectrum for the as machined sample (top) shows prominent peaks originating from $\mathrm{Nb_2O_5}$, and is commonly observed on untreated niobium surfaces. 
In contrast, the spectrum for the EDM-finished sample is completely different.
As mentioned in the main text, an EDM-finished cavity exhibits an order of magnitude smaller $Q_{\rm int}$ than cavities fabricated by machining alone,
and its temperature dependence differs significantly (blue symbols in Fig.~\ref{fig Q-T, all trace}).
This behavior most likely originates from the EDM-induced formation of a low-$T_c$ material on the cavity surface.
It is widely known that EDM forms a recast layer tens of micrometers thick at the surface, which contains carbon and hydrogen contamination from insulating oil used in EDM.
This is consistent with the absence of metallic-Nb related peaks in the EDM-finished sample.

After $900\,^{\circ}\mathrm{C}$ baking following BCP flush, $\mathrm{Nb_2O_5}$ ratio decreases compared to the untreated niobium sample surface.
The composition ratio of $\mathrm{Nb_2O_5}$, estimated from the peak area of XPS spectrum, decreases from 87.8\,\% to 57.5\,\% after a water-buffered BCP flush (wBCP-f) and to 61.0\,\% after conventional BCP flush (BCP-f).
$600\,^{\circ}\mathrm{C}$ annealing further reduces the $\mathrm{Nb_2O_5}$ ratio to 40.8\,\%.
Meanwhile, NbO ratio increases from 0.83\,\% (as machined) to 12.3\,\% (wBCP-f) and 11.9\,\% (BCP-f) after $900\,^{\circ}\mathrm{C}$ baking, eventually reaching 43.7\,\% after $600\,^{\circ}\mathrm{C}$ annealing.

It should be noted that metallic-Nb peaks around 202\,eV become undetectable after $600\,^{\circ}\mathrm{C}$ annealing, suggesting the presence of a thick NbO layer just beneath the outermost $\mathrm{Nb_2O_5}$ layer.
This likely helps to suppress the regrowth of $\mathrm{Nb_2O_5}$ upon air exposure and these observations are consistent with an earlier report by Yu~\cite{Yu_Vacuum2022}.

\section{Cavity performance details}\label{appendix_Alltrace}

The temperature dependence of $Q_{\rm int}$ for all cavities measured in this work is shown in Figure~\ref{fig Q-T, all trace}, and the extracted parameters are listed in Table~\ref{appendix: tab-value}.

\begin{table*}[t!]
  \begin{ruledtabular}
  \begin{tabular}{cccccccccc}
    &
    & 
    &
    \multicolumn{3}{c}{$T<20\,\mathrm{mK}$} &
    \multicolumn{2}{c}{$T=1.2\,\mathrm{K}$} &
    \multicolumn{2}{c}{TLS fit}
    \\
    Cavity &
    Treatment &
    Cooldown &
    \textrm{$Q_{\rm int}$}  &
    \textrm{$R_{s}$}  &
    \textrm{$\tau_{\rm int}$} &
    \textrm{$Q_{\rm int}$ } &
    \textrm{$R_{s}$} &
    \textrm{$F \delta^0_{\rm TLS}$} &
    \textrm{$Q_{\rm res}$} 
    \\
    & 
    &
    Cycles &
    \textrm{($\times 10^6$)} &  
    \textrm{[n$\Omega$]} &
    \textrm{[ms]} &
    \textrm{($\times 10^6$)} &  
    \textrm{[n$\Omega$]} & 
    & 
    \\  
  \colrule
  A1 & No treatment & & 32.9 & 2000 & 0.969 & 60.9 & 1080 & $1.43\times10^{-8}$ & $4.23\times10^{7}$  \\  
  & BCP + HT  & 1 & 358 & 185 & 10.5 & 12.3 & 5410 & $1.15\times10^{-9}$ & $5.33\times10^{8}$  \\
  &  & 2 & 331 & 200 & 9.67 & 12.7 & 5210 & $1.85\times10^{-10}$ & 3.49$\times10^{8}$  \\
  & BCP + HT + wBCP-f & & 455 & 146 & 13.3 & - & - & $2.93\times10^{-10}$ & $5.26\times10^{9}$  \\
  & BCP + HT + wBCP-f$\,\times\,2$ &  & 1150 & 57.6 & 33.5 & 4590 & 14.5 & $6.39\times10^{-10}$ & $3.05\times10^{9}$  \\
  B1 & BCP + HT & & 150 & 441 & 4.4 & 17.6 & 3750 & - & -  \\
  & BCP + HT + wBCP-f & & 663 & 98 & 19.4 & - & - & $9.68\times10^{-10}$ & $1.81\times10^{9}$  \\
  & BCP + HT + wBCP-f + BCP-f & 1 & 871 & 76.0 & 25.4 & 4590 & 14.4 & $9.93\times10^{-10}$ & $6.50\times10^{9}$ \\
  &  & 2 & 828 & 80.0 & 24.2 & 4110 & 16.1 & $1.06\times10^{-9}$ & $5.27\times10^{9}$ \\
  &  & 3 & 538 & 123 & 15.7 & 1190 & 55.5 & $1.19\times10^{-9}$ & $1.46\times10^{9}$ \\
  A2 & No treatment & & 48.7 & 1350 & 1.43 & 97.4 & 674 & $4.31\times10^{-10}$ & $4.98\times10^{7}$  \\
  & BCP + HT + BCP-f & 1 & 917 & 72.3 & 26.7 & 1890 & 35.0 & $9.62\times10^{-10}$ & $4.92\times10^{9}$  \\
  &  & 2 & 784 & 84.5 & 22.9 & 1870 & 35.3 & $1.22\times10^{-9}$ & $4.60\times10^{9}$ \\
  & BCP + HT + BCP-f + $460\,^{\circ}\mathrm{C}$\,-\,3h & 1 & 1220 & 54.3 & 35.6 & - & - & $4.85\times10^{-10}$ & $2.90\times10^{9}$  \\
  &  & 2 & 1180 & 56.0 & 34.5 & 2390 & 27.8 & $4.98\times10^{-10}$ & $3.22\times10^{9}$  \\
  &  & 3 & 1250 & 52.9 & 36.6 & - & - & $4.67\times10^{-10}$ & $2.86\times10^{9}$  \\
  &  & 4 & 1240 & 53.6 & 36.0 & - & - & $4.90\times10^{-10}$ & $2.80\times10^{9}$  \\
  B2 & No treatment & & 6.31 & 10400 & 0.185 & - & - & $1.32\times10^{-8}$ & $6.84\times10^{6}$  \\
  & BCP + HT + BCP-f & 1 & 1140 & 58.0 & 33.3 & 5570 & 11.9 & $6.13\times10^{-9}$ & $3.56\times10^{9}$  \\
  &  & 2 & 286 & 232 & 8.32 & 377 & 10.9 & $1.61\times10^{-9}$ & $5.08\times10^{8}$ \\
  & BCP + HT + BCP-f + $600\,^{\circ}\mathrm{C}$\,-\,3h & 1 & 3030 & 21.9 & 88.3 & 3190 & 20.8 & $2.10\times10^{-10}$ & $7.41\times10^{9}$  \\
  &  & 2 & 2640 & 25.1 & 77.0 & 26.1 & 74.2 & $1.82\times10^{-10}$ & $4.77\times10^{9}$ \\
  \end{tabular}
  \end{ruledtabular}
  \caption{%
  Summary of the internal quality factors $Q_{\rm int}$, internal lifetimes, residual resistances, and loss tangents of 3D coaxial cavities with various surface treatments.
  (BCP): 100$\,\mu\mathrm{m}$ BCP, (HT): 900\,$^\circ\mathrm{C}$\,-\,3\,h high-temperature baking, (BCP-f): BCP flush with standard BCP mixture, (wBCP-f): water-buffered BCP flush.
  The parameters for TLS fitting ($F \delta^0_{\rm TLS}$ and $Q_{\rm res}$) are extracted from the average photon number dependence of $Q_{\rm int}$ at base temperature $T<20\,\mathrm{mK}$. 
  }
  \label{appendix: tab-value}
\end{table*}

\bibliographystyle{unsrt}
\bibliography{High-Q_Nb}

\begin{thebibliography}{10}

\bibitem{Paik_PRL2011}
Hanhee Paik, D.~I. Schuster, Lev~S. Bishop, G.~Kirchmair, G.~Catelani, A.~P. Sears, B.~R. Johnson, M.~J. Reagor, L.~Frunzio, L.~I. Glazman, S.~M. Girvin, M.~H. Devoret, and R.~J. Schoelkopf.
\newblock {Observation of High Coherence in Josephson Junction Qubits Measured in a Three-Dimensional Circuit QED Architecture}.
\newblock {\em Phys. Rev. Lett.}, 107:240501, Dec 2011.

\bibitem{Reagor_PRB2016}
Matthew Reagor, Wolfgang Pfaff, Christopher Axline, Reinier~W. Heeres, Nissim Ofek, Katrina Sliwa, Eric Holland, Chen Wang, Jacob Blumoff, Kevin Chou, Michael~J. Hatridge, Luigi Frunzio, Michel~H. Devoret, Liang Jiang, and Robert~J. Schoelkopf.
\newblock {Quantum memory with millisecond coherence in circuit QED}.
\newblock {\em Phys. Rev. B}, 94:014506, Jul 2016.

\bibitem{Ofek_Nature2016}
Nissim Ofek, Andrei Petrenko, Reinier Heeres, Philip Reinhold, Zaki Leghtas, Brian Vlastakis, Yehan Liu, Luigi Frunzio, Steven~M Girvin, Liang Jiang, et~al.
\newblock {Extending the lifetime of a quantum bit with error correction in superconducting circuits}.
\newblock {\em Nature}, 536(7617):441--445, 2016.

\bibitem{Gao_Nature2019}
Yvonne~Y Gao, Brian~J Lester, Kevin~S Chou, Luigi Frunzio, Michel~H Devoret, Liang Jiang, SM~Girvin, and Robert~J Schoelkopf.
\newblock {Entanglement of bosonic modes through an engineered exchange interaction}.
\newblock {\em Nature}, 566(7745):509--512, 2019.

\bibitem{Sivak_Nature2023GKP}
Volodymyr~V Sivak, Alec Eickbusch, Baptiste Royer, Shraddha Singh, Ioannis Tsioutsios, Suhas Ganjam, Alessandro Miano, BL~Brock, AZ~Ding, Luigi Frunzio, et~al.
\newblock {Real-time quantum error correction beyond break-even}.
\newblock {\em Nature}, 616(7955):50--55, 2023.

\bibitem{Ni_Nature2023Bin11}
Zhongchu Ni, Sai Li, Xiaowei Deng, Yanyan Cai, Libo Zhang, Weiting Wang, Zhen-Biao Yang, Haifeng Yu, Fei Yan, Song Liu, et~al.
\newblock {Beating the break-even point with a discrete-variable-encoded logical qubit}.
\newblock {\em Nature}, 616(7955):56--60, 2023.

\bibitem{Checchin_PhysRevApplied2022}
Mattia Checchin, Daniil Frolov, Andrei Lunin, Anna Grassellino, and Alexander Romanenko.
\newblock {Measurement of the low-temperature loss tangent of high-resistivity silicon using a high-Q superconducting resonator}.
\newblock {\em Physical Review Applied}, 18(3):034013, 2022.

\bibitem{Read_PhysRevApplied2023}
Alexander~P Read, Benjamin~J Chapman, Chan~U Lei, Jacob~C Curtis, Suhas Ganjam, Lev Krayzman, Luigi Frunzio, and Robert~J Schoelkopf.
\newblock {Precision measurement of the microwave dielectric loss of sapphire in the quantum regime with parts-per-billion sensitivity}.
\newblock {\em Physical Review Applied}, 19(3):034064, 2023.

\bibitem{Romanenko_PRL2023}
A~Romanenko, R~Harnik, A~Grassellino, R~Pilipenko, Y~Pischalnikov, Z~Liu, OS~Melnychuk, B~Giaccone, O~Pronitchev, T~Khabiboulline, et~al.
\newblock {Search for dark photons with superconducting radio frequency cavities}.
\newblock {\em Physical review letters}, 130(26):261801, 2023.

\bibitem{Agrawal_PRL2024}
Ankur Agrawal, Akash~V Dixit, Tanay Roy, Srivatsan Chakram, Kevin He, Ravi~K Naik, David~I Schuster, and Aaron Chou.
\newblock {Stimulated emission of signal photons from dark matter waves}.
\newblock {\em Physical review letters}, 132(14):140801, 2024.

\bibitem{Nakazono_cQED-Darkmatter2025}
K~Nakazono, S~Chen, H~Fukuda, Y~Iiyama, T~Inada, T~Moroi, T~Nitta, A~Noguchi, R~Sawada, S~Shirai, et~al.
\newblock {Search for Dark Photon Dark Matter of a Mass around $\ensuremath 36.1\,{\mu}$eV Using a Frequency-tunable Cavity Controlled through a Coupled Superconducting Qubit}.
\newblock {\em arXiv preprint arXiv:2505.15619}, 2025.

\bibitem{Kudra_APL2020}
M.~Kudra, J.~Biznárová, A.~Fadavi~Roudsari, J.~J. Burnett, D.~Niepce, S.~Gasparinetti, B.~Wickman, and P.~Delsing.
\newblock {{High quality three-dimensional aluminum microwave cavities}}.
\newblock {\em Applied Physics Letters}, 117(7):070601, 08 2020.

\bibitem{Kubo:IPAC2014-WEPRI022}
T.~Kubo, Y.~Ajima, H.~Inoue, K.~Umemori, Y.~Watanabe, and M.~Yamanaka.
\newblock {{I}n{-}house {P}roduction of a {L}arge{-G}rain {S}ingle{-C}ell {C}avity at {C}avity {F}abrication {F}acility and {R}esults of {P}erformance {T}ests}.
\newblock In {\em Proc. 5th International Particle Accelerator Conference (IPAC'14), Dresden, Germany, June 15-20, 2014}, number~5 in International Particle Accelerator Conference, pages 2519--2521, Geneva, Switzerland, July 2014. JACoW.
\newblock https://doi.org/10.18429/JACoW-IPAC2014-WEPRI022.

\bibitem{Romanenko_PRL2017}
A.~Romanenko and D.~I. Schuster.
\newblock {Understanding Quality Factor Degradation in Superconducting Niobium Cavities at Low Microwave Field Amplitudes}.
\newblock {\em Phys. Rev. Lett.}, 119:264801, Dec 2017.

\bibitem{Romanenko_PRAppl2020}
A.~Romanenko, R.~Pilipenko, S.~Zorzetti, D.~Frolov, M.~Awida, S.~Belomestnykh, S.~Posen, and A.~Grassellino.
\newblock {Three-Dimensional Superconducting Resonators at ${T}<20$ mK with Photon Lifetimes up to $\ensuremath{\tau}=2$ s}.
\newblock {\em Phys. Rev. Appl.}, 13:034032, Mar 2020.

\bibitem{HshGL}
Mark~K. Transtrum, Gianluigi Catelani, and James~P. Sethna.
\newblock {Superheating field of superconductors within Ginzburg-Landau theory}.
\newblock {\em Phys. Rev. B}, 83:094505, Mar 2011.

\bibitem{HshModerate}
F.~Pei-Jen Lin and A.~Gurevich.
\newblock {Effect of impurities on the superheating field of type-II superconductors}.
\newblock {\em Phys. Rev. B}, 85:054513, Feb 2012.

\bibitem{HshDirty}
Takayuki Kubo.
\newblock {Superfluid flow in disordered superconductors with Dynes pair-breaking scattering: Depairing current, kinetic inductance, and superheating field}.
\newblock {\em Phys. Rev. Res.}, 2:033203, Aug 2020.

\bibitem{Padamsee}
Hasan Padamsee.
\newblock {50 years of success for SRF accelerators—a review}.
\newblock {\em Superconductor Science and Technology}, 30(5):053003, apr 2017.

\bibitem{FE}
Bianca Giaccone, Martina Martinello, Paolo Berrutti, Oleksandr Melnychuk, Dmitri~A. Sergatskov, Anna Grassellino, Dan Gonnella, Marc Ross, Marc Doleans, and John~F. Zasadzinski.
\newblock {Field emission mitigation studies in the SLAC Linac Coherent Light Source II superconducting rf cavities via in situ plasma processing}.
\newblock {\em Phys. Rev. Accel. Beams}, 24:022002, Feb 2021.

\bibitem{Cenni_FE}
Hiroshi Sakai, Enrico Cenni, Kazuhiro Enami, Takaaki Furuya, Masaru Sawamura, Kenji Shinoe, and Kensei Umemori.
\newblock {Field emission studies in vertical test and during cryomodule operation using precise x-ray mapping system}.
\newblock {\em Phys. Rev. Accel. Beams}, 22:022002, Feb 2019.

\bibitem{hydride}
Akshay Murthy Daniel Bafia Evguenia Karapetrova Martina Martinello Jaeyel Lee Anna~Grassellino Zuhawn~Sung, Arely~Cano and Alexander Romanenko.
\newblock {Direct observation of nanometer size hydride precipitations in superconducting niobium}.
\newblock {\em Sci. Rep.}, 14:26916, 2024.

\bibitem{Qdesease}
J.~M. Cavedon M. Juillard A. Godin C. Henriot Ph. Leconte H. Safa A.~Veyssiere B.~Aune, B.~Bonin and C.~Zylberajch.
\newblock {Degradation of Niobium Superconducting RF Cavities During Cooling Time}.
\newblock In {\em Proc. the Linear Accelerator Conference 1990, Albuquerque, New Mexico, USA, 1990}, page 253, Geneva, Switzerland, 1990. JACoW.

\bibitem{Qdesease_Saito}
K.~Saito and P.~Kneisel.
\newblock {Q-Degradation in High Purity Niobium Cavities-Dependence on Temperature and RRR-Value}.
\newblock In {\em Proc. the Third European Particle Accelerator Conference, Berlin, Germany, 1992}, page 1231, Geneva, Switzerland, 1992. JACoW.

\bibitem{KyotoCamera}
Yoshihisa Iwashita, Yujiro Tajima, and Hitoshi Hayano.
\newblock {Development of high resolution camera for observations of superconducting cavities}.
\newblock {\em Phys. Rev. ST Accel. Beams}, 11:093501, Sep 2008.

\bibitem{Ge_2011}
M~Ge, G~Wu, D~Burk, J~Ozelis, E~Harms, D~Sergatskov, D~Hicks, and L~D Cooley.
\newblock {Routine characterization of 3D profiles of SRF cavity defects using replica techniques}.
\newblock {\em Superconductor Science and Technology}, 24(3):035002, dec 2010.

\bibitem{Romanenko_APL2014}
A.~Romanenko, A.~Grassellino, A.~C. Crawford, D.~A. Sergatskov, and O.~Melnychuk.
\newblock {Ultra-high quality factors in superconducting niobium cavities in ambient magnetic fields up to 190 mG}.
\newblock {\em Applied Physics Letters}, 105(23):234103, 12 2014.

\bibitem{PhysRevAccelBeams.19.082001}
Shichun Huang, Takayuki Kubo, and R.~L. Geng.
\newblock {Dependence of trapped-flux-induced surface resistance of a large-grain Nb superconducting radio-frequency cavity on spatial temperature gradient during cooldown through ${T}_{c}$}.
\newblock {\em Phys. Rev. Accel. Beams}, 19:082001, Aug 2016.

\bibitem{Milul_PRXQuantum2023}
Ofir Milul, Barkay Guttel, Uri Goldblatt, Sergey Hazanov, Lalit~M. Joshi, Daniel Chausovsky, Nitzan Kahn, Engin \ifmmode~\mbox{\c{C}}\else \c{C}\fi{}ifty\"urek, Fabien Lafont, and Serge Rosenblum.
\newblock {Superconducting Cavity Qubit with Tens of Milliseconds Single-Photon Coherence Time}.
\newblock {\em PRX Quantum}, 4:030336, Sep 2023.

\bibitem{Oriani_arXiv2024}
Andrew~E. Oriani, Fang Zhao, Tanay Roy, Alexander Anferov, Kevin He, Ankur Agrawal, Riju Banerjee, Srivatsan Chakram, and David~I. Schuster.
\newblock {Niobium coaxial cavities with internal quality factors exceeding 1.5 billion for circuit quantum electrodynamics}.
\newblock {\em arXiv preprint arXiv:2403.00286}, 2024.

\bibitem{Posen_2020}
S.~Posen, A.~Romanenko, A.~Grassellino, O.S. Melnychuk, and D.A. Sergatskov.
\newblock {Ultralow Surface Resistance via Vacuum Heat Treatment of Superconducting Radio-Frequency Cavities}.
\newblock {\em Phys. Rev. Appl.}, 13:014024, Jan 2020.

\bibitem{Ito_PTEP2021}
H~Ito, H~Araki, K~Takahashi, and K~Umemori.
\newblock {Influence of furnace baking on Q–E behavior of superconducting accelerating cavities}.
\newblock {\em Progress of Theoretical and Experimental Physics}, 2021(7):071G01, 04 2021.

\bibitem{CFF}
H.~Inoue T. Kubo T. Saeki Y. Watanabe S.~Yamaguchi M.~Yamanaka, Y.~Ajima.
\newblock {Cavity Fabrication Study in CFF at KEK}.
\newblock In {\em Proceedings of SRF2013, Paris, France, 2013}, number~16 in International Conference on RF Superconductivity, pages 821--824, Geneva, Switzerland, 2013. JACoW.

\bibitem{Kalboussi_PhysRevApplied2025}
Y.~Kalboussi, I.~Curci, F.~Miserque, D.~Troadec, N.~Brun, M.~Walls, G.~Jullien, F.~Eozenou, M.~Baudrier, L.~Maurice, Q.~Bertrand, P.~Sahuquet, and T.~Proslier.
\newblock {Crystallinity in niobium oxides: A pathway to mitigate two-level-system defects in niobium three-dimensional resonators for quantum applications}.
\newblock {\em Phys. Rev. Appl.}, 23:044023, Apr 2025.

\bibitem{Probst_RSI2015}
S.~Probst, F.~B. Song, P.~A. Bushev, A.~V. Ustinov, and M.~Weides.
\newblock {Efficient and robust analysis of complex scattering data under noise in microwave resonators}.
\newblock {\em Review of Scientific Instruments}, 86(2):024706, 02 2015.

\bibitem{Gao_APL2008}
Jiansong Gao, Miguel Daal, Anastasios Vayonakis, Shwetank Kumar, Jonas Zmuidzinas, Bernard Sadoulet, Benjamin~A. Mazin, Peter~K. Day, and Henry~G. Leduc.
\newblock {Experimental evidence for a surface distribution of two-level systems in superconducting lithographed microwave resonators}.
\newblock {\em Appl. Phys. Lett.}, 15(92):152505, 2008.

\bibitem{Wang_APL2009}
H.~Wang, M.~Hofheinz, J.~Wenner, M.~Ansmann, R.~C. Bialczak, M.~Lenander, Erik Lucero, M.~Neeley, A.~D. O’Connell, D.~Sank, M.~Weides, A.~N. Cleland, and John~M. Martinis.
\newblock {Improving the coherence time of superconducting coplanar resonators}.
\newblock {\em Applied Physics Letters}, 95(23):233508, 12 2009.

\bibitem{Muller_RevProgPhys2019}
Clemens Müller, Jared~H Cole, and Jürgen Lisenfeld.
\newblock {Towards understanding two-level-systems in amorphous solids: insights from quantum circuits}.
\newblock {\em Reports on Progress in Physics}, 82(12):124501, oct 2019.

\bibitem{McRae_RSI2020}
C.~R.~H. McRae, H.~Wang, J.~Gao, M.~R. Vissers, T.~Brecht, A.~Dunsworth, D.~P. Pappas, and J.~Mutus.
\newblock {Materials loss measurements using superconducting microwave resonators}.
\newblock {\em Review of Scientific Instruments}, 91(9):091101, 09 2020.

\bibitem{Lei_APL2020}
Chan~U Lei, Lev Krayzman, Suhas Ganjam, Luigi Frunzio, and Robert~J. Schoelkopf.
\newblock {High coherence superconducting microwave cavities with indium bump bonding}.
\newblock {\em Applied Physics Letters}, 116(15):154002, 04 2020.

\bibitem{Altoe_PRXQuantum2022}
M.~Virginia~P. Alto\'e, Archan Banerjee, Cassidy Berk, Ahmed Hajr, Adam Schwartzberg, Chengyu Song, Mohammed Alghadeer, Shaul Aloni, Michael~J. Elowson, John~Mark Kreikebaum, Ed~K. Wong, Sin\'ead~M. Griffin, Saleem Rao, Alexander Weber-Bargioni, Andrew~M. Minor, David~I. Santiago, Stefano Cabrini, Irfan Siddiqi, and D.~Frank Ogletree.
\newblock {Localization and Mitigation of Loss in Niobium Superconducting Circuits}.
\newblock {\em PRX Quantum}, 3:020312, Apr 2022.

\bibitem{Gurevich_2017}
Alex Gurevich and Takayuki Kubo.
\newblock {Surface impedance and optimum surface resistance of a superconductor with an imperfect surface}.
\newblock {\em Phys. Rev. B}, 96:184515, Nov 2017.

\bibitem{Kubo_2022}
Takayuki Kubo.
\newblock {Effects of Nonmagnetic Impurities and Subgap States on the Kinetic Inductance, Complex Conductivity, Quality Factor, and Depairing Current Density}.
\newblock {\em Phys. Rev. Appl.}, 17:014018, Jan 2022.

\bibitem{Heidler_PRAppl2021}
Paul Heidler, Christian M.~F. Schneider, Katja Kustura, Carlos Gonzalez-Ballestero, Oriol Romero-Isart, and Gerhard Kirchmair.
\newblock {Non-Markovian Effects of Two-Level Systems in a Niobium Coaxial Resonator with a Single-Photon Lifetime of 10 milliseconds}.
\newblock {\em Phys. Rev. Appl.}, 16:034024, Sep 2021.

\bibitem{Kalboussi_APL2024}
Y.~Kalboussi, B.~Delatte, S.~Bira, K.~Dembele, X.~Li, F.~Miserque, N.~Brun, M.~Walls, J.~L. Maurice, D.~Dragoe, J.~Leroy, D.~Longuevergne, A.~Gentils, S.~Jublot-Leclerc, G.~Jullien, F.~Eozenou, M.~Baudrier, L.~Maurice, and T.~Proslier.
\newblock {Reducing two-level systems dissipations in 3D superconducting niobium resonators by atomic layer deposition and high temperature heat treatment}.
\newblock {\em Applied Physics Letters}, 124(13):134001, 03 2024.

\bibitem{Bafia_PhysRevApplied2024}
D.~Bafia, A.~Murthy, A.~Grassellino, and A.~Romanenko.
\newblock {Oxygen vacancies in niobium pentoxide as a source of two-level system losses in superconducting niobium}.
\newblock {\em Phys. Rev. Appl.}, 22:024035, Aug 2024.

\bibitem{Yu_Vacuum2022}
Mingming Yu, Guo Pu, Yi~Xue, Sishu Wang, Sheng Chen, Yihan Wang, Li~Yang, Zhijun Wang, Tongtong Zhu, Teng Tan, Yuan He, Shichun Huang, and Kun Zhang.
\newblock {The oxidation behaviors of high-purity niobium for superconducting radio-frequency cavity application in vacuum heat treatment}.
\newblock {\em Vacuum}, 203:111258, 2022.

\bibitem{Verjauw_PhysRevApplied2021}
J.~Verjauw, A.~Poto\ifmmode~\check{c}\else \v{c}\fi{}nik, M.~Mongillo, R.~Acharya, F.~Mohiyaddin, G.~Simion, A.~Pacco, Ts. Ivanov, D.~Wan, A.~Vanleenhove, L.~Souriau, J.~Jussot, A.~Thiam, J.~Swerts, X.~Piao, S.~Couet, M.~Heyns, B.~Govoreanu, and I.~Radu.
\newblock {Investigation of Microwave Loss Induced by Oxide Regrowth in High-Q Niobium Resonators}.
\newblock {\em Phys. Rev. Appl.}, 16:014018, Jul 2021.

\bibitem{1993_Kneisel}
N.~Pupeter E.~Mahner, P.~Kneisel and G.~Muller.
\newblock {Effect of Chemical Polishing on the Electron Field Emission of Niobium Samples and Cavities}.
\newblock In {\em Proc. the Sixth Workshop on RF Superconductivity, CEBAF, Newport News, Virginia, USA, 1993}, page 1085, Geneva, Switzerland, 1993. JACoW.

\bibitem{1989_Saito}
T.~Furuya S. Mitsunobu S. Noguchi K. Hosoyama T. Nakazato T. Tajima K. Asano K. Inoue Y. Iino H.~Nomura K.~Saito, Y.~Kojima and K.~Takeuchi.
\newblock {R \& D of Superconducting Cavities at KEK}.
\newblock In {\em Proc. the Fourth Workshop on RF Superconductivity, KEK, Tsukuba, Japan, 1989}, page 635, Geneva, Switzerland, 1989. JACoW.

\bibitem{Kneisel_flux}
B.~Piosczyk, P.~Kneisel, O.~Stoltz, and J.~Halbritter.
\newblock {Investigations of Additional Losses in Superconducting Niobium Cavities Due to Frozen-In Flux}.
\newblock {\em IEEE Transactions on Nuclear Science}, 20(3):108--112, 1973.

\bibitem{Posen_flux}
S.~Posen, M.~Checchin, A.~C. Crawford, A.~Grassellino, M.~Martinello, O.~S. Melnychuk, A.~Romanenko, D.~A. Sergatskov, and Y.~Trenikhina.
\newblock {Efficient expulsion of magnetic flux in superconducting radiofrequency cavities for high Q applications}.
\newblock {\em Journal of Applied Physics}, 119(21):213903, 06 2016.

\bibitem{2021_Ooi}
S.~Ooi, M.~Tachiki, T.~Konomi, T.~Kubo, A.~Kikuchi, S.~Arisawa, H.~Ito, and K.~Umemori.
\newblock {Observation of intermediate mixed state in high-purity cavity-grade Nb by magneto-optical imaging}.
\newblock {\em Phys. Rev. B}, 104:064504, Aug 2021.

\bibitem{2025_Ooi}
S.~Ooi, M.~Tachiki, T.~Mochiku, H.~Ito, T.~Kubo, A.~Kikuchi, S.~Arisawa, and K.~Umemori.
\newblock {Dynamical visualization of attractively interacting single vortices in type-II/1 superconducting Nb by magneto-optical imaging}.
\newblock {\em Phys. Rev. B}, 111:094519, Mar 2025.

\bibitem{Konomi}
T.~Konomi et~al.
\newblock {{T}rial of {N}itrogen {I}nfusion and {N}itrogen {D}oping by {U}sing {J-PARC} {F}urnace}.
\newblock In {\em Proc. of International Conference on RF Superconductivity (SRF'17), Lanzhou, China, July 17-21, 2017}, number~18 in International Conference on RF Superconductivity, pages 775--778, Geneva, Switzerland, Jan. 2018. JACoW.
\newblock https://doi.org/10.18429/JACoW-SRF2017-THPB021.

\bibitem{2020_Marc}
Marc Wenskat, Christopher Bate, Arti Dangwal~Pandey, Arno Jeromin, Thomas~F Keller, Jens Knobloch, Julia Köszegi, Felix Kramer, Oliver Kugeler, Satish Kulkarni, Detlef Reschke, Jörn Schaffran, Guilherme Dalla Lana~Semione, Sven Sievers, Lea Steder, Andreas Stierle, and Nicholas Walker.
\newblock {Nitrogen infusion R\&D at DESY a case study on cavity cut-outs}.
\newblock {\em Superconductor Science and Technology}, 33(11):115017, oct 2020.

\bibitem{EXFEL}
W.~Singer, A.~Brinkmann, R.~Brinkmann, J.~Iversen, A.~Matheisen, W.-D. Moeller, A.~Navitski, D.~Reschke, J.~Schaffran, A.~Sulimov, N.~Walker, H.~Weise, P.~Michelato, L.~Monaco, C.~Pagani, and M.~Wiencek.
\newblock {Production of superconducting 1.3-GHz cavities for the European X-ray Free Electron Laser}.
\newblock {\em Phys. Rev. Accel. Beams}, 19:092001, Sep 2016.

\bibitem{bake}
K.~Twarowski P. Schmuser D. Bloess E. Haebel E. Chiaveri J.-M. Tessier H. Preis H. Wenninger H.~Safa L.~Lilje, D.~Reschke and J.-P. Charrier.
\newblock {Electropolishing and in-situ Baking of 1.3 GHz Niobium Cavities}.
\newblock In {\em Proc. the Nineth Workshop on RF Superconductivity, La Fonda Hotel, Santa Fe, New Mexico, USA, 1999}, page~74, Geneva, Switzerland, 1999. JACoW.

\bibitem{Bafia_SRF2019}
D.~Bafia, A.~Grassellino, O.S. Melnychuk, A.S. Romanenko, Z-H. Sung, and J.~Zasadzinski.
\newblock {Gradients of 50 MV/m in TESLA Shaped Cavities via Modified Low Temperature Bake}.
\newblock In {\em Proc. SRF'19}, number~19 in International Conference on RF Superconductivity, pages 586--591. JACoW Publishing, Geneva, Switzerland, aug 2019.
\newblock https://doi.org/10.18429/JACoW-SRF2019-TUP061.

\bibitem{Ciovati_bake}
Gianluigi Ciovati.
\newblock {Improved oxygen diffusion model to explain the effect of low-temperature baking on high field losses in niobium superconducting cavities}.
\newblock {\em Applied Physics Letters}, 89(2):022507, 07 2006.

\bibitem{Lechner_2024}
E.~M. Lechner, J.~W. Angle, A.~D. Palczewski, F.~A. Stevie, M.~J. Kelley, and C.~E. Reece.
\newblock {Oxide dissolution and oxygen diffusion scenarios in niobium and implications on the Bean–Livingston barrier in superconducting cavities}.
\newblock {\em Journal of Applied Physics}, 135(13):133902, 04 2024.

\bibitem{Kubo_2017}
Takayuki Kubo.
\newblock {Multilayer coating for higher accelerating fields in superconducting radio-frequency cavities: a review of theoretical aspects}.
\newblock {\em Superconductor Science and Technology}, 30(2):023001, dec 2016.

\bibitem{Wave_2019}
Vudtiwat Ngampruetikorn and J.~A. Sauls.
\newblock {Effect of inhomogeneous surface disorder on the superheating field of superconducting RF cavities}.
\newblock {\em Phys. Rev. Res.}, 1:012015, Aug 2019.

\bibitem{Kubo_2021}
Takayuki Kubo.
\newblock {Superheating fields of semi-infinite superconductors and layered superconductors in the diffusive limit: structural optimization based on the microscopic theory}.
\newblock {\em Superconductor Science and Technology}, 34(4):045006, mar 2021.

\bibitem{Sha_2022}
Song Jin Ji-Yuan Zhai Zheng-Hui Mi Bai-Qi Liu Chao Dong Fei-Si He Rui Ge Liang-Rui Sun Shi-Ao~Zheng Peng~Sha, Wei-Min~Pan and Ling-Xi Ye.
\newblock {Ultrahigh accelerating gradient and quality factor of CEPC 650 MHz superconducting radio-frequency cavity}.
\newblock {\em Nuclear Science and Techniques}, 33:125, 2022.

\bibitem{Kubo_Gurevich_2019}
Takayuki Kubo and Alex Gurevich.
\newblock {Field-dependent nonlinear surface resistance and its optimization by surface nanostructuring in superconductors}.
\newblock {\em Phys. Rev. B}, 100:064522, Aug 2019.

\bibitem{HPR}
W.~Hartung C. Hauviller-W. Weingarten P.~Bosland Ph.~Bernard, D.~Bloess and J.~M. Martignac.
\newblock {Superconducting Niobium Sputter-Coated Copper Cavities at 1500 MHz}.
\newblock In {\em Proc. the Fifth Workshop on RF Superconductivity, DESY, Hamburg, Germany, 1991}, page 487, Geneva, Switzerland, 1991. JACoW.

\bibitem{HPR_Saito}
K.~Kurosawa P. Kneisel-S. Noguchi E.Kako M.Ono T.~Shishido K.~Saito, H.~Miwa and T.~Suzuki.
\newblock {Study of Ultra-clean Surface for Niobium SC Cavities}.
\newblock In {\em Proc. the Sixth Workshop on RF Superconductivity, CEBAF, Newport News, Virginia, USA, 1993}, page 1151, Geneva, Switzerland, 1993. JACoW.

\bibitem{LG4X-V2}
Julian~Andreas Hochhaus and Hideki Nakajima.
\newblock {LG4X-V2 (2.3.1)}.
\newblock \url{https://doi.org/10.5281/zenodo.16418092}.

\bibitem{Prudnikava_SST2024}
Alena Prudnikava, Yegor Tamashevich, Anna Makarova, Dmitry Smirnov, and Jens Knobloch.
\newblock {In-situ synchrotron x-ray photoelectron spectroscopy study of medium-temperature baking of niobium for SRF application}.
\newblock {\em Superconductor Science and Technology}, 37(7):075007, 2024.

\end{thebibliography}

\end{document}